\documentclass[a4paper]{article}

\usepackage{jheppub}

\usepackage[utf8x]{inputenc}
\usepackage[english]{babel}
\usepackage{graphicx}
\usepackage{wrapfig}
\usepackage{amsmath}
\usepackage{url}
\usepackage{amsfonts}
\usepackage{subfig}
\usepackage{alltt}
\usepackage{fancyhdr}
\usepackage{tensor}
\usepackage{amssymb}
\usepackage{bbm}
\usepackage{color}
\usepackage[yyyymmdd]{datetime}
\usepackage{float}
\usepackage{amsfonts}
	\makeatletter
	\def\@fpheader{\phantom{Prepared for submission to JHEP}}
	\makeatother

\DeclareMathOperator{\Tr}{Tr}

\DeclareSymbolFont{bbold}{U}{bbold}{m}{n}
\DeclareSymbolFontAlphabet{\mathbbold}{bbold}

\numberwithin{equation}{section}

\title{Analysis of constraints and their algebra in bimetric theory} 

\author{S.~F. Hassan}
\author{and Anders Lundkvist}

\affiliation{Department of Physics \& The Oskar Klein Centre, \\
	Stockholm University, AlbaNova University Centre, SE-106 91 Stockholm, Sweden}

\emailAdd{fawad@fysik.su.se}
\emailAdd{anders.lundkvist@fysik.su.se}

\abstract{We perform a canonical analysis of the bimetric theory in
  the metric formulation, computing the constraints and their algebra
  explicitly. In particular, we compute a secondary constraint, that
  has been argued to exist earlier, and show that it has the correct
  form to eliminate the ghost. We also identify a set of four first
  class constraints that generate the algebra of general
  covariance. The covariance algebra naturally determines a spacetime
  metric for the theory. However, in bimetric theory, this metric is
  not unique but depends on how the first class constraints are
  identified.}
\begin{document}
\toccontinuoustrue
\maketitle

\newcommand{\f}{\phi}
\newcommand{\h}{\ensuremath{ \; \! {}^3 \: \! \!  h}}
\newcommand{\R}{\ensuremath{ \; \! {}^3 \: \! \! \! R}}
\newcommand{\diff}{\mathrm{d}}
\newcommand{\ghat}{\hat{\gamma}}
\newcommand{\fhat}{\hat{\phi}}
\newcommand{\bal}{\begin{align}}
\newcommand{\eal}{\end{align}}    
\newcommand{\C}{\ensuremath{\mathcal{C}_{(2)}}}
\newcommand{\p}{\partial}

\section{Introduction} \label{intro}

Bimetric theory is a theory of the gravitational metric interacting
with another spin-2 field. The study of such a theory can be motivated
by comparing general relativity with Maxwell's theory and noting that
they are the simplest possible theories of a spin-2 and a spin-1
field, respectively. However, while Maxwell's theory works perfectly
well at the classical level, and to a large extent even at the quantum
level, in reality it is not a standalone theory. Rather, it is
embeded in the $SU(2)_W\times U(1)_Y$ electroweak theory. In fact, all
fields in the standard model appear in multiplets, which is crucial
for the consistency and observational viability of the theory. This
feature of the standard model suggests the possibility that general
relativity could arise from a setup including more spin-2
fields. However, constructing theories of multiple spin-2 fields is
not straightforward, since such theories have redundant field
components, including ghost modes \cite{Boulware-Deser}, which must be
eliminated by appropriate constraints present in the theory. 

In this paper we carry out an analysis of constraints in the ghost
free bimetric theory \cite{bimetric,HK}, which is formulated in terms
of two spin-2 fields $g_{\mu\nu}$ and $f_{\mu\nu}$. We compute the
constraints explicitly, in particular a secondary constraint that was
argued to exist in \cite{conf}, showing that it has the correct form
to eliminate the ghost. We also identify a set of four first class
constraints that generate the algebra of general covariance.  From the
work of Hojman, Kucha\v{r} and Teitelboim (HKT) it known that one can
read off a metric from this algebra \cite{hkt}. We show that this
metric is not unique, but depends on what combination of constraints
are chosen to form a first class set.

Ghost free massive gravity \cite{dRG,dRGT,HR1}, which can be obtained
from bimetric theory by freezing the dynamics of $f_{\mu\nu}$, has
fewer constraints, due to the lack of general covariance. These were
computed in \cite{Hassan1106,fra}, in \cite{conf}, and subsequently
in \cite{pilo1,pilo2,HSvS-stuckelberg,deffayet1,deffayet2,deffayet3,deser,
  Kugo:2014hja}. In bimetric theory, of the six constraints needed,
five were obtained in \cite{bimetric} and the sixth one was argued to
exist in \cite{conf}. More work on this has been carried out in
\cite{kluson1301,kluson1303,kluson1307,krasnov,alexandrov, soloviev1,
   soloviev2,Bernard:2015uic}. In this paper we address the computation
of the sixth constraint and of the algebra of constraints, as detailed
below. 

The paper is organized as follows. In the remainder of this section we
review the known results on bimetric constraints and then summarize
the results of the present paper. Section \ref{bimetric} gives a
review of the ghost free bimetric theory and presents an outline of the
calculations in our reduced phase space approach.  In section
\ref{biseccon}, the secondary constraint of the bimetric theory is
evaluated explicitly and shown to have the required form. Section
\ref{algebra-chapter} deals with finding the first class constraints
in bimetric theory and identifying the HKT metric from the general
covariance algebra. In appendix \ref{det_Lagrange}, all Lagrange
multipliers of the theory are determined at the linear level. Detailed
derivations of the Poisson brackets of constraints are relegated to
appendix \ref{appendix_brackets}.

\subsection{Background and summary of results}
We consider a gravitational metric $g_{\mu\nu}$ with an
Einstein-Hilbert action, interacting with another spin-2 field
represented by a symmetric tensor $f_{\mu\nu}$, via a potential
$V(g^{-1}f)$. Then the necessary condition for the absence of ghosts
\cite{Creminelli:2005qk} determines the form of the potential
\cite{dRG,dRGT,HR1} and also dictates the dynamics of $f_{\mu\nu}$ to
be given by an Einstein-Hilbert term \cite{bimetric}, leading to a
bimetric action,  
\begin{align}
{\cal S} &= \int \mathrm{d}^4x \left(M_g^2 
\sqrt{|\det g|} R^{(g)}+ M_f^2 \sqrt{|\det f|} R^{(f)} + 2m^4 
 \sqrt{|\det g|} V(g^{-1} f)\right) 
\text{.}
\end{align}
To eliminate the unwanted field components, the theory must contain
constraints, four of which are associated with general covariance.  

In \cite{HK} it was shown that general covariance and the reality of
the bimetric equations restrict the two metrics such that their null
cones always have a nonvanishing intersection. An implication is that
one can always find coordinate systems in which the two metrics admit
simultaneous $3+1$ decompositions in terms of their respective lapses,
shifts and spatial 3-metrics,
\begin{align}
&N \equiv \left(-g^{00} \right)^{-1/2} \text{,} \qquad N_i \equiv
  g_{0i} \text{,} \qquad \gamma_{ij} \equiv g_{ij} \text{,} 
\label{ADMvariables} \\
&L \equiv \left(-f^{00} \right)^{-1/2} \text{,} \qquad L_i \equiv
f_{0 i} \text{,} \qquad \f_{ij}\equiv f_{ij} \text{.} \label{fADMvariables}
\end{align}
This decomposition, used in \cite{fra,bimetric}, is convenient for
analyzing the dynamical content of the theory and covers the most
general case. 

Because of the Einstein-Hilbert terms, only time derivatives of
$\gamma_{ij}$ and $\f_{ij}$ appear in the action, hence there are 12
potentially propagating degrees of freedom. The 8 components of the
lapses and shifts are nondynamical. On general grounds, 4 of the
potentially dynamical modes (and their conjugate momenta) can be
eliminated by gauge fixing general covariance and the associated
constraints. Of the remaining 8 degrees of freedom, 7 are physical,
corresponding to a massless and a massive spin-2 field, and the remaining
mode is the Boulware-Deser ghost. Two constraints are needed to eliminate the
ghost field and its conjugate momentum. Hence we need 6 constraints in
total (along with the equations to determine the lapses and shifts). 

After some manipulations, one set of the shift variables, say, the
$N_i$, can be eliminated using their equations of motion, leading to 
a Hamiltonian with $N, L$ and $L^i$ as Lagrange multipliers, 
\cite{bimetric,fra},   
\begin{equation}
  \mathcal{H}=
  -\left(N \mathcal{C} + L \tilde{R}^0 + L^i \tilde{R}_i\right).
\end{equation} 
This yields the first five constraints, $\mathcal{C} = 0$,
$\tilde{R}^0 = 0$, and $\tilde{R}_i = 0$.
Since these equations arise from an action principle, they are valid
at all times, in particular $\dot{\mathcal{C}} \approx 0$, where the
symbol $\approx$ is used for weak equalities, i.e. equalities that
hold on the surface of the constraints. This equation can be used 
to obtain a new condition easily expressed in terms of the Poisson
bracket,
\begin{equation} \label{intro_biC2}
\C \equiv \frac{\diff\mathcal{C}}{\diff t}=\{\mathcal{C},H\} \approx 0\,.
\end{equation}
This will be the last constraint needed, provided it has the correct
structure. We will see that $\C \approx 0$ also ensures the preservation of 
$\tilde{R}^0$, and that $\tilde{R}_i$ are automatically preserved in
time.

Let us first review the situation in massive gravity where
$f_{\mu\nu}$ is a fixed nondynamical metric and two constraints are
needed to eliminate the ghost. In this case $\mathcal{C}$ is the only
constraint obtained directly from the Hamiltonian and it preservation
in time gives,
\begin{equation} \label{intro_massiveC2}
\C=\frac{\partial\mathcal{C}}{\partial t}+\{\mathcal{C},H\} \approx 0\,.  
\end{equation}
The $\partial\mathcal{C}/\partial t$ term accounts for the explicit
time dependence via the background $f_{\mu\nu}$ which is no longer
computed by the Poisson bracket. It was shown in \cite{conf} that in
massive gravity $\{\mathcal{C}(x),\mathcal{C}(y) \}\approx 0$, hence,
$\C$ is independent of $N$ and provides the needed secondary
constraint. The explicit form of $\C$, valid for time independent
$f_{\mu\nu}$, was computed. Furthermore, it was argued that since
these properties also hold in bimetric theory, equation
\eqref{intro_biC2} is a constraint in that case as well. The massive
gravity constraints have also been obtained in
\cite{pilo1,pilo2,HSvS-stuckelberg} and further confirmed in
\cite{deffayet1,deffayet2,deffayet3,deser} without a 3+1
decomposition. They were used for the analysis of classical solutions
in \cite{volkov1402,volkov1404}.  Note that $\C$ in massive gravity 
depends on the lapse $L$ and shift $L_i$ of the nondynamical
$f_{\mu\nu}$ \cite{conf}, which is not a problem since these are
prespecified functions. 

{\bf Constraints in Bimetric Theory:} It was pointed out in
\cite{kluson1301,kluson1303} that also in bimetric theory, $\C$, given
by \eqref{intro_biC2}, depends on the lapse $L$ and its spatial
derivative $\partial_i L$. It was argued that then $\C \approx 0$ is
not a constraint since it would determine $L$ rather than eliminate
the momentum conjugate to the ghost field. In this paper, we show that
this is not the case. We perform a canonical analysis of the bimetric
theory in the metric formulation, computing all constraints and their
stability conditions. In particular $\C$ is calculated explicitly and
shown to contain $L$ only as an overall factor. All terms involving
$L^i$ and $\partial_i L$ are shown to vanish. Hence $\C\approx 0$ is a
valid secondary constraint, ensuring the absence of the Boulware-Deser
ghost. It is also shown that the stability condition
$\dot{\mathcal{C}}_{(2)} \approx 0$ can be solved for $N$ in terms of
$L$. Then, $L$ and $L^i$ can be determined through gauge fixing
general coordinate transformations, allowing us to express all
Lagrange multipliers in terms of the dynamical variables. This is
explicitly worked out at the linear level in appendix
\ref{det_Lagrange}. The explicit expressions obtained are also needed 
for solving the initial value problem.

The canonical analysis of bimetric theory has also been performed in
\cite{krasnov,alexandrov}, in chiral and tetrad variables,
respectively. They conclude that the theory does have the constraints
to eliminate the ghost and propagates the appropriate seven degrees of
freedom. However, due to the different formalisms used, it not trivial
to directly compare their expression for the secondary constraints
with the ones obtained here.

{\bf{First class constraints and the HKT metric:}} GR has a
Hamiltonian $\mathcal{H}_{GR} = -N^\mu R_\mu$, leading to constraints
$R_0\approx 0$, $R_i\approx 0$, which in turn satisfy the general
covariance algebra, 
\begin{align} \label{intro_algebra}
\{R^0(x) , R^0(y) \} &= - \left[ R^i(x) \frac{\partial}{\partial x^i} \delta^3(x-y) - R^i(y) \frac{\partial}{\partial y^i} \delta^3(x-y) \right] \text{,} \notag \\
\{R^0(x) , R_i(y) \} &= - R^0(y) \frac{\partial}{\partial x^i} \delta^3(x-y) \text{,} \\
\{R_i(x) , R_j(y) \} &= - \left[ R_j(x) \frac{\partial}{\partial x^i} \delta^3(x-y) - R_i(y) \frac{\partial}{\partial y^j} \delta^3(x-y) \right] \text{.} \notag
\end{align}
In fact, this is a feature of any covariant theory \cite{Teitelboim}. The algebra
contains both $R_i$ and $R^i$, hence it explicitly depends on a
3-metric that relates the two. In GR this is the spatial metric
$g_{ij}$. 

In \cite{hkt} Hojman, Kucha\v{r} and Teitelboim conjectured that this
observation can be used to identify a 3-metric in theories which have
general covariance but where a unique metric is not {\it a priori}
specified. Obviously, bimetric theory is such an example. The theory
admits two independent matter sectors each coupled minimally to one of
the metrics \cite{bimetric,HSS}. Therefore, restricting to one matter 
sector, the covariance algebra leads to either $g_{ij}$ or $f_{ij}$
as the HKT metric. However, the full theory can also contain 
other combinations of $g$ and $f$ as effective metrics, and {\it a
  priori} it is not obvious if a preferred combination exists and how
this is compatible with the matter sectors.

In \cite{krasnov}, the constraint algebra was studied in a chiral
formulation of the bimetric theory. The metric identified using the
HKT-conjecture is a complicated function of the variables used and
does not coincide with either $g_{\mu\nu}$ or $f_{\mu\nu}$.  
Some aspects of the bimetric constraint algebra have been considered
in \cite{soloviev1, soloviev2}, in
\cite{kluson1301,kluson1307,kluson1303}, as well as in 
\cite{alexandrov}, but without addressing the HKT metric.  

Having obtained the constraints, in this paper we also compute the
constraint algebra. We identify a set of 4 first class constraints
among the six outlined above, and show that they satisfy the standard
algebra of general covariance. The first class constraints can be
easily identified by imposing the second class constraints and the
stability condition $\dot{\cal C}_2 \approx 0$, to bring the bimetric
Hamiltonian to the form $\mathcal{H} = -L^\mu R_\mu$. The HKT metric
appearing in the covariance algebra is then found to be
$f_{ij}$. However, using an alternative set of variables modifies the
first class constraints and leads to $g_{ij}$ as the HKT metric. The
explanation is that a set of first class constraints can be identified
among all the constraints in different ways, each choice leading to a
different HKT metric. The possibility of other choices, beyond the two
mentioned above, is also discussed. It is possible that the metric
identified in \cite{krasnov} corresponds to one such choice. All this
is consistent with matter couplings, since their structure is similar
to the gravitational sector.

In this paper the constraints have mainly been used to argue for the
absence of ghosts and the correct counting of degrees of freedom.  In
addition, they are also essential for determining valid initial data
that will be evolved by the remaining dynamical equations.  However,
for bimetric theory to be fully consistent, it is also necessary that the
dynamical equations produce causal evolution.  Some aspects of this
problem have been investigated in
\cite{Izumi:2013poa,Camanho:2016opx,HK,Bellazzini:2017fep,
  Hinterbichler:2017qyt,Bonifacio:2017nnt,Hinterbichler:2017qcl}.

\section{Review of constraints in bimetric gravity} \label{bimetric}

In this section we review the derivation of the constraints in
bimetric theory following \cite{bimetric,conf}. 

\subsection{The Hamiltonian formulation of bimetric theory} \label{biADM}

The most general ghost free bimetric action for spin-2 fields
$g_{\mu \nu}$ and $f_{\mu \nu}$ is \cite{bimetric},
\begin{align} \label{bi-action}
{\cal S} &= \int \mathrm{d}^4x \left(M_g^2 
\sqrt{|\det g|} R^{(g)}+ M_f^2 \sqrt{|\det f|} R^{(f)} + 2m^4 
 \sqrt{|\det g|} \sum_{n=0}^4 \beta_n e_n ( \sqrt{g^{-1} f} )\right) 
\text{,}
\end{align}
where $R^{(g)}$ and $R^{(f)}$ are the respective Ricci scalars, $M_g$
and $M_f$ are the corresponding Planck masses, and $\beta_n$ are five
free parameters. The interactions between $g_{\mu\nu}$ and
$f_{\mu\nu}$ involve the square root matrix $S=\sqrt{g^{-1}f}$, which
is a specific root of the matrix $g^{-1}f$ to be specified below. The
$e_k (S)$ are elementary symmetric polynomials of the eigenvalues of
$S$ given by (with $\Tr(S)=[S]$),
\begin{align}
e_0(S) &= 1 \text{,} \quad 
e_1(S) = [S] \text{,} \quad 
e_2(S) = \frac{1}{2} \left( [S]^2 - [S^2] \right) \text{,} \quad
e_3(S) = \frac{1}{6} \left( [S]^3 - 3 [S] [S^2] + 2 [S^3] \right)
\text{,} \notag \\
e_4(S) &= \frac{1}{24} \left( [S]^4 - 6 [S]^2 [S^2] + 3 [S^2]^2 + 8
[S] [S^3] - 6 [S^4] \right)
\text{.} 
\label{esps}
\end{align}
For a $4 \times 4$ matrix, $e_4(S)\equiv\det S$ and $e_k(S)\equiv 0$
for $k>4$. Note that $\det(1+S)=\sum_{n=0}^4 e_n(S)$, so the potential
$\sum_{n=0}^4\beta_n e_n (S)$ is a deformation of $\det(1+S)$. Each
term has the property,       
\begin{equation} \label{bi-symmetry}
\sqrt{|\det g|}\,  e_n ( \sqrt{g^{-1} f} ) =
\sqrt{|\det f|}\, e_{4-n} ( \sqrt{f^{-1} g} ) 
\text{,} 
\end{equation}
hence, the action \eqref{bi-action} retains its form under the
interchange of $g_{\mu\nu}$ and $f_{\mu\nu}$ \cite{bimetric}. The
theory can be easily generalized to any dimension $d$.

By definition, the matrix $S=\sqrt{g^{-1}f}$ must satisfy 
\cite{dRGT,HR1},       
\begin{equation}
{\left( \sqrt{g^{-1} f} \right)^\mu}_\lambda {\left( \sqrt{g^{-1} f}
  \right)^\lambda}_\nu = g^{\mu \lambda} f_{\lambda \nu} \text{.} 
\end{equation}
This does not specify the square root uniquely and may even
admit nonreal solutions. Obviously, as long as $S$ is not uniquely
specified, the action \eqref{bi-action} remains ill defined. This
problem has a natural resolution as follows (see \cite{HK} for
details). In order for the action to be invariant under coordinate
transformations, $S^\mu_{~\nu}$ must transform as a $(1,1)$ tensor and
only the {\it principal root} has this property.\footnote{All
  non-principal roots cease to transform as $(1,1)$ tensors whenever
  eigenvalues of $g^{-1}f$ belonging to distinct Jordan blocks happen
  to coincide.}  Furthermore, to obtain real equations of motion, $g$
and $f$ must be restricted such that a {\it real} principal root
always exists. This restriction turns out to imply that the null cones
of the metrics $g$ and $f$ have a nonvanishing intersection, so that
they admit common spacelike and timelike directions \cite{HK}. Thus,
the requirements of reality and general covariance specify the theory
uniquely by restricting $S$ to the {\it real principal} root of
$g^{-1} f$. The same requirements also ensure that all allowed $g$
and $f$ admit simultaneous proper 3+1 decompositions.

Therefore, without loss of generality, we can use the 3+1 metric
decompositions \eqref{ADMvariables} and \eqref{fADMvariables} in the
bimetric action. As in GR, the action will only contain time
derivatives of the 3-metrics $\gamma_{ij}$ and $\f_{ij}$. Denoting the
corresponding canonical momenta by $\pi^{ij}$ and $p^{ij}$, the
Lagrangian can be readily expressed in terms of the phase space
variables, as in \cite{ADM},
\begin{align} \label{biLag_intermed}
\mathcal{L} =\left(\pi^{ij} \dot{\gamma}_{ij} + N R^{0(g)} + N^i R_i^{(g)}\right) 
&+ \left(p^{ij} \dot{\f}_{ij} + L R^{0(f)} + L^i R_i^{(f)} \right) \nonumber\\
& + 2m^4 N
\sqrt{\det \gamma} \sum_{n=0}^3 \beta_n e_n(\sqrt{g^{-1} f}) \text{.}  
\end{align}
The first line simply contains the $3+1$ decomposition of the
Einstein-Hilbert terms for $g$ and $f$. Here, $N^i = \gamma^{ij} N_j$
and $L^i=\f^{ij} L_j$, where $\gamma^{ij}$ and $\f^{ij}$ are the
inverses of $\gamma_{ij}$ and $\f_{ij}$, respectively. As in general
relativity, $R^{0(g)}$, $R_i^{(g)}$, $R^{0(f)}$ and $R_i^{(f)}$ are
given by,    
\begin{align}
R^{0(g)} &= M_g^2 \sqrt{\det\gamma}\, \R^{(g)} + \frac{1}{M_g^2
  \sqrt{\det \gamma}} \left(\tfrac{1}{2} \pi^2 - \pi^{ij} \pi_{ij}
\right) \text{,} \quad
R_i^{(g)} = 2 \sqrt{\det\gamma}\, \gamma_{ij} \nabla_k 
\left(\frac{\pi^{jk}}{\sqrt{\det \gamma}} \right) \text{,} 
\\
R^{0 (f)} &= M_f^2 \sqrt{\det\f}\, \R^{(f)} + \frac{1}{M_f^2
\sqrt{\det\f}}\left(\tfrac{1}{2}p^2-p^{ij}p_{ij}\right) \text{,}    
\quad 
R_i^{(f)} = 2 \sqrt{\det \f}\, \f_{ij} \bar{\nabla}_k 
\left(\frac{p^{jk}}{\sqrt{\det \f}} \right) \text{,} 
\end{align} 
where $\nabla_i$ and $\bar{\nabla}_i$ are the covariant derivatives
compatible with $\gamma_{ij}$ and $\f_{ij}$, respectively. 

In \eqref{biLag_intermed}, the lapses and shifts, $N$, $L$, $N^i$ and
$L^i$, appear without time derivatives. The dynamical variables are
$\gamma_{ij}$, $\f_{ij}$, $\pi^{ij}$ and $p^{ij}$ with a total of 24
(phase space) components, corresponding to 12 potentially propagating
degrees of freedom, including potential ghost modes\footnote{In this
  usage, a degree of freedom consists of two phase space variables,
  {\it i.e.}, the field and its conjugate momentum.}. The theory has
enough symmetries and constraints to eliminate the ghost and reduce
the number of propagating modes to seven, corresponding to a massless
and a massive spin-2 field.  However, since the potential term in
\eqref{biLag_intermed} is nonlinear in the lapses and shifts, some
manipulations are needed to make the constraints manifest. This is
outlined in subsection \ref{primconst}, where the constraint that
eliminates the ghost field is obtained. A second constraint that
eliminates the canonical momentum of the ghost field is computed in
section \ref{biseccon}.

\subsection{Outline of the analysis}

To obtain the constraint equations, we work in a reduced phase space.
Before getting into the details, it is worthwhile to briefly outline
our procedure and show, in particular, that it is equivalent to the
more elaborate formalism of
\cite{kluson1301,kluson1303,kluson1307,pilo1,pilo2} involving an
enlarged phase space. In the next subsection we will see that the
bimetric Lagrangian has the general form $L(q_m, \dot{q}_m, Q_r,
n_i)$, depending on a set of dynamical variables $q_m(t)$ and two sets
of nondynamical variables $Q_r(t)$ and $n_i(t)$. Here, for simplicity
we ignore the space dependence of fields and gauge invariances. The
nonvanishing canonical momenta are $p^m=\p L/\p\dot{q}_m$ conjugate to
$q_m$. The Lagrangian equations of motion are,
\begin{align}
\dot{p}^m=\frac{\p L}{\p q_m}\,,\quad \frac{\p L}{\p Q_m}=0\,,
\quad\frac{\p L}{\p n_i}=0\,.
\end{align}
These contain all the information about dynamics and constraints.  To
completely disentangle the two, we rewrite the action in terms of
phase space variables using the Hamiltonian $H=p^m\dot{q}_m-L$,
\begin{align}
S=\int dt\left[p^m\dot{q}_m-H(p,q,Q,n)\right] 
\end{align}
Then, on varying with respect to $q_m, p^m, Q_r$, and $n_i$,  one
obtains the equivalent set of equations,
\begin{align} \label{PBeqs}
  \dot{p}^m=-\frac{\p H}{\p q_m}\,,\quad
  \dot{q}_m=\frac{\p H}{\p p^m}\,,\quad
  \quad \frac{\p H}{\p Q_m}=0\,,\quad \frac{\p H}{\p n_i}=0\,.
\end{align}
It turns out that the $n_i$ equations determine $n_i=n_i(q,p)$,
independent of $Q_r$, which can be used to eliminate the 
$n_i$ from the action. Although now the action develops an extra
dependence on $p^m$ and $q_m$ through the $n_i$, in practice, 
variations of $H$ can be computed at fixed $n_i$, since,
\begin{align}
\delta H = \delta\big|_{n_i} H + \frac{\p H}{\p
  n_i}\Big|_{q_m,p^m,Q_r}\delta n_i
\end{align}
and the last term vanishes for the solutions of the $n_i$ equations.
Hence, effectively, $H=H(q, p, Q)$.  In bimetric theory, the $Q_r$
appear as Lagrange multipliers and the Hamiltonian has the form
$H={\cal H}_r(p,q) Q_r$. It follows that $\p{\cal H}_r/\p n_i=0$,
either being proportional to the $n_i$ equations of motion or because
some ${\cal H}_r$ did not contain $n_i$ to begin with. In terms of
the Poisson brackets defined with respect to the conjugate pairs
 $(q_m, p^m)$, the above equations can be recast as,
\begin{align}
\dot{p}^m=\{p^m, H\}\,,\qquad \dot{q}^m=\{q^m, H\}\,,
\quad {\cal H}_r(p,q)=0\,.
\end{align}
The constraints on ${\cal H}_r$ can be solved to eliminate an equal
number of dynamical variables. 

In bimetric theory the above equations hide one more constraint that
can be extracted as follows. Since the constraints are obtained from
the action principle, they hold at all times, in particular, $d{\cal
  H}_r/dt=0$. One of the constraints, say ${\cal H}_1\equiv C$
is special. Since $C$ varies only with $p^m$ and $q_m$, we get 
$dC/dt=\{C, H\}$, using \eqref{PBeqs}. Hence, $\dot{C}=0$ combined
with \eqref{PBeqs} implies,
\begin{align}
C_2\equiv\{C, H\} =0\,.
\end{align}
This turns out to be independent of the $Q_r$ and thus is a new
constraint. One can also compute $dC_2/dt= \{C_2, H\}=0$, but this
equation contains some of the $Q_r$ and is not a constraint on the
$q_m$ and $p^m$. The remaining ${\cal H}_r$ do not lead to new
constraints in this way. Note that, although expressed in terms of 
Poisson brackets, the above equations are derived from the action
principle. 

Let's now compare the above analysis to the formalism of constrained
Hamiltonian dynamics applied to the same problem
\cite{kluson1301,kluson1303,kluson1307,pilo1,pilo2}. This is based on
an enlarged phase space spanned by variables $q_m, Q_r$ and their
conjugate momenta $p^m, P^r$, where the primary constraints $P^r=0$
are enforced through Lagrange multipliers $u_r$. The relevant
Hamiltonian is,
\begin{align}
H'= H(p,q,Q)+ P^r u_r\,.
\end{align}
We can easily extend this to also include the nondynamical variables 
$n_i$ and their conjugate momenta, say $m^i=0$. Equations of
motion are now given in terms of the Poisson brackets with $H'$ on the
enlarged phase space. One can easily check that $\dot{p}^m = \{p^m,
H'\}'$ and $\dot{q}^m=\{q^m, H'\}'$ reduce to the corresponding
equations in the reduced formalism given above. The preservation of
the primary constraints in time then gives $\dot{P^r}=\{P^r,H'\}'=0$
which reproduce the constraints ${\cal H}_i=0$. If $n_i$ are included,
their equations of motion will also arise in this way. The new
equations are $\dot{Q_r}=\{Q_r,H'\}'=u_r$ which reflect the fake
dynamics of the $Q_r$. Finally, the preservation of $C={\cal H}_1$ in
time leads to the same constraint $C_2\equiv\dot{C}=\{C, H'\}' =0$ as
above, and the $\dot{C}_2=0$ equation is unchanged as can be checked by
using the properties of the other constraints. Hence working in the
reduced phase space is equivalent to the framework of the constrained
Hamiltonian dynamics. 

\subsection{Constraints from lapse and shift equations} \label{primconst}

In order to deal with the square root matrix in the bimetric action
and show that the equations for the eight lapse and shift variables
encode five constraints, it is convenient to introduce the shift
like variables $n^i$ through
\cite{Hassan1106,fra,dRGT},\footnote{\label{fn-symmetry} Due to the
  interchange symmetry \eqref{bi-symmetry}, the following analysis
  also holds with the roles of $g$ and $f$ interchanged.}
\begin{equation} \label{ndef}
N^i - L^i = L n^i + N {D^i}_j n^j \text{.}
\end{equation}
The $3\times 3$ matrix $D$ can be obtained explicitly as a solution to
the equation \cite{fra}, 
\begin{equation} \label{Ddef}
\sqrt{x} D = \sqrt{ \left( \gamma^{-1} - D n n^T D^T \right) \f } \text{\,.}
\end{equation}
Such a solution always exists \cite{HKS}, but is not needed here. The
quantity $x$ stands for,  
\begin{equation} \label{xdef}
x \equiv 1 - n^i \f_{ij} n^j \text{.}
\end{equation}
From equation \eqref{Ddef}, it follows that $D$ has the property,
\begin{equation} \label{Dfsymmetry}
\f_{ik} {D^k}_j = \f_{jk} {D^k}_i \text{.}
\end{equation} 
Now, on eliminating $N^i$ in favor of $n^i$, the action corresponding
to \eqref{biLag_intermed} becomes, \cite{bimetric}
\begin{equation} \label{bi-Lagrangian}
\mathcal{S} = \int \mathrm{d}^4x \left(\pi^{ij} 
\partial_t \gamma_{ij} + p^{ij} \partial_t
\f_{ij} + L^i \tilde{R}_i + L \tilde{R}^0 + N \mathcal{C}\right)
 \text{\,,} 
\end{equation}
where $N$, $L$ and $L^i$ appear as five Lagrange multipliers and
we have defined, 
\begin{align} \label{R0tilde_def}
\tilde{R}^0&=R^{0 (f)}+n^i R_i^{(g)}+2m^4 \sqrt{\det \gamma}U'\text{,}  
\\
\label{Ritilde_def}
\tilde{R}_i &=  R_i^{(g)} + R_i^{(f)} \text{,}
\\
\label{bi-C}
\mathcal{C}\, &= R^{0 (g)} + R_i^{(g)} {D^i}_j n^j + 2m^4 
\sqrt{\det \gamma} V \text{,}
\end{align}
with,
\begin{equation} \label{U'-def}
\sqrt{\det \gamma} U' = \sqrt{\det \gamma} U + \beta_4 \sqrt{\det \f} \text{.}
\end{equation}
$U$ and $V$ contain the parameters $\beta_n$, and are given by,
\begin{align} \label{Udef}
U \equiv & \beta_1 \sqrt{x} + \beta_2 \left[ x\, e_1(D) + n^i \f_{ij}
  {D^j}_k n^k \right] 
\notag \\
&\qquad\qquad + \beta_3 \big[ \sqrt{x} \left( e_1(D) n^i \f_{ij} {D^j}_k n^k -
{D^i}_k n^k \f_{ij}{D^j}_l n^l\right)+x^{3/2} e_2(D)\big]\text{,} 
\\
\label{Vdef}
V \equiv & \beta_0 + \beta_1 \sqrt{x}\, e_1(D) + \beta_2 x\, e_2(D) +
\beta_3 x^{3/2}\, e_3(D) \text{.} 
\end{align}
The $e_n(D)$ are defined similar to \eqref{esps}, but now 
$e_3(D)=\det D$, since $D$ is a $3\times 3$ matrix.  

In the action \eqref{bi-Lagrangian}, the equations of motion for $N$,
$L$ and $L^i$ are $\mathcal{C}=0$, $\tilde{R}^0=0$ and
$\tilde{R}_i=0$. However, $\mathcal{C}$ and $\tilde{R}^0$ depend on
the redefined shifts $n^i$. Therefore, in order for them to impose
constraints on the dynamical variables, it is necessary that the
equations for $n^i$ are independent of the Lagrange multipliers. Only
then will $\mathcal{C}$, $\tilde{R}^0$ and $\tilde{R}_i$ impose five
constraints on the dynamical variables.  Indeed, the $n^k$ equations
of motion are,
\begin{equation} \label{dL/dn}
\frac{\partial \mathcal{L}}{\partial n^k} = L \frac{\partial
  \tilde{R}^0}{\partial n^k} + N \frac{\partial \mathcal{C}}{\partial
  n^k} = 0 \text{.} 
\end{equation}
But, from equation \eqref{Ddef}, it follows that \cite{fra},
\begin{align} \label{dR0/dn}
\frac{\partial \tilde{R}^0}{\partial n^k} = \mathcal{C}_k \text{,} \qquad\quad
\frac{\partial \mathcal{C}}{\partial n^k} = \mathcal{C}_i
\frac{\partial ({D^i}_j n^j)}{\partial n^k} \text{,} 
\end{align}
where
\begin{align} \label{biC_i}
\mathcal{C}_i = R_i^{(g)} - 2 & m^4 \sqrt{\det \gamma} \frac{n^l
\f_{lj}}{\sqrt{x}} \Big[ \beta_1 {\delta^j}_i + \beta_2 \sqrt{x}
\left( {\delta^j}_i {D^m}_m - {D^j}_i \right) \notag \\ 
&+ \beta_3 x \left(\tfrac{1}{2} {\delta^j}_i \left( {D^m}_m {D^n}_n -
{D^m}_n {D^n}_m \right) + {D^j}_m {D^m}_i - {D^j}_i {D^m}_m
\right) \Big] \text{.} 
\end{align}
Inserting \eqref{dR0/dn} into \eqref{dL/dn} yields
the equation, 
\begin{equation}
\mathcal{C}_i \left[ L {\delta^i}_k + N \frac{\partial ({D^i}_j
    n^j)}{\partial n^k} \right] = 0 \text{.} 
\end{equation}
The expression inside the square brackets is the Jacobian matrix
$\partial N^i/\partial n^k$ of \eqref{ndef}, and is invertible. Hence,
the $n^i$ equations of motion imply,\footnote{Since ${\partial
    \mathcal{L}}/{\partial n^k}=({\partial\mathcal{L}}/{\partial N^i})
  ({\partial N^i}/{\partial n^k})$, \eqref{neqom} are simply the
  equations of motion for $N^i$ expressed in terms of $n^i$.} 
\begin{equation} \label{neqom}
\mathcal{C}_i = 0 \text{.}
\end{equation}
These are independent of $N$, $L$ and $L^i$ and can in principle be
solved for the $n^i$. In the $\beta_1$-model (i.e. when
$\beta_2=\beta_3=0$) it is easy to obtain the explicit solution, while
perturbative solutions can be constructed for general $\beta_n$
\cite{fra}. When these solutions are inserted in the equations of
motion for $N$, $L$ and $L^i$, one obtains five constraints on
$\gamma_{ij}$, $\f_{ij}$, $\pi^{ij}$ and $p^{ij}$,
\begin{align}
\mathcal{C}=0\,, \qquad \tilde{R}^0=0\,, \qquad \tilde{R}_i=0\,.
\label{fiveConst}
\end{align}
$\tilde{R}^0$ and $\tilde{R}_i$ can be associated with general
covariance while $\mathcal{C}=0$ eliminates the BD ghost field. A new
constraint is needed to remove the momentum conjugate to the ghost and
is reviewed below.

\subsection{The existence of the extra secondary constraint} 
\label{secconstraint} 

The five constraints \eqref{fiveConst} obtained from an action
principle are valid at all times, in particular, $\dot{\mathcal{C}}
\equiv d{\mathcal{C}}/dt=0$ holds. Contrast this with the Hamiltonian
formulation where such conditions must be imposed additionally,
leading to new constraints (see for example,
\cite{Dirac1964}). However, even here recasting in Hamiltonian form,
$\dot{\mathcal{C}}=\{{\mathcal{C}}, H\}$, is useful as it requires
eliminating $\dot\gamma_{ij}$, etc., using the dynamical
equations. Then $\dot{\mathcal{C}}=0$ extracts a new secondary
constraint from the dynamical equations.  Specifically, if $\C \equiv
\{{\mathcal{C}}, H\}\approx 0$ does not involve Lagrange multipliers,
it will be a new constraint on the dynamical variables (otherwise it
could be solved for a Lagrange multiplier).\footnote{In the terminology
  of \cite{Dirac1964}, the vanishing of the momenta conjugate to the
  lapses and shifts are the primary constraints, and the validity of
  these at all times leads to \eqref{fiveConst} as secondary
  constraints. The validity of $\mathcal{C}=0$ at all times gives the
  new secondary constraint $\C=0$.}

The explicit expression for $\C$ in massive gravity with a fixed
$f_{\mu\nu}$ was obtained in \cite{conf}. It is independent of $N$,
but contains $L$ and $L^i$. This is not an issue since in massive
gravity these are prespecified functions and $\C\approx 0$ remains a
constraint on the dynamical variables. However, in bimetric theory,
where $f_{\mu\nu}$ is not prespecified, such a dependence is
problematic. Indeed, in \cite{kluson1301,kluson1303} it was argued
that, in bimetric theory, $\C$ depends on $L$ and $\partial_i L$ (as
one may expect from \eqref{C2compbracket}) and hence is no longer a
constraint on dynamical variables. Note that, in principle, $\C$ could
still turn out to be a constraint since $L$ and $L_i$ can be
determined in terms of dynamical variables through gauge fixing
general covariance. 

In the remainder of this section, we review the argument in
\cite{conf} that a secondary constraint exists in both theories. Then,
in section \ref{biseccon} we will go beyond \cite{conf} to explicitly
evaluate $\C$ in bimetric theory and show that its vanishing does not
depend on $L$ and $L_i$, hence it is a genuine constraint. We also
show that no further constraints arise since $\dot{\mathcal{C}}_{(2)}
\approx 0$ determines the Lagrange multiplier $N$. $L$ and $L_i$ are
determined through gauge fixing, similarly to what is done in GR. At
the linear level, this is explicitly worked out in appendix
\ref{det_Lagrange}.

In bimetric theory $\mathcal{C}$ depends on the dynamical
variables, hence using Poisson brackets one has,
\begin{equation} \label{C2def}
\mathcal{C}_{(2)}(x) = \frac{d}{dt} \mathcal{C}(x) =
\{\mathcal{C}(x),H \} \approx 0 \text{.}
\end{equation}
The Hamiltonian can be read off from the action in
\eqref{bi-Lagrangian} as,  
\begin{equation} \label{bi-Hamiltonian}
H = - \int \diff^3 y \left( N(y) \mathcal{C}(y) + L(y) \tilde{R}^0(y)
+ L^i(y) \tilde{R}_i(y) \right) \text{.} 
\end{equation}
Then it follows that,
\begin{equation} \label{C2compbracket}
\mathcal{C}_{(2)}(x) \approx - \int \mathrm{d}^3y
\left(N(y)\{\mathcal{C}(x), \mathcal{C}(y) \}  + L^i(y)  \{\mathcal{C}(x), 
  \tilde{R}_i(y) \} + L(y) \{ \mathcal{C}(x), \tilde{R}^0(y) \}
  \right) \approx 0  \text{,} 
\end{equation}
where the Poisson bracket is defined by
\begin{align} \label{poisson}
\{A,B \} &= \int \mathrm{d}^3 z \left( \frac{\delta A}{\delta
  \gamma_{mn}(z)} \frac{\delta B}{\delta \pi^{mn}(z)} - 
\frac{\delta A}{\delta \pi^{mn}(z)} 
\frac{\delta B}{\delta \gamma_{mn}(z)} \right) \notag \\
&+ \int \mathrm{d}^3 z \left(\frac{\delta A}{\delta \f_{mn}(z)} 
\frac{\delta B}{\delta p^{mn}(z)} - \frac{\delta A}{\delta p^{mn}(z)} 
\frac{\delta B}{\delta \f_{mn}(z)} \right) \text{.}
\end{align} 
Note that if $\{\mathcal{C}(x),\mathcal{C}(y)\} \not \approx 0$, then
\eqref{C2compbracket} could determine $N$, rather than impose a
constraint on the dynamical variables, as suggested in
\cite{kluson1301}. Hence, for $\C$ to be a constraint, it is necessary
that $\{ \mathcal{C}(x),\mathcal{C}(y) \} \approx 0$, while in
bimetric theory \eqref{C2compbracket} must also be independent of $L$
and $L_i$. This bracket is evaluated using \eqref{poisson}, but since
$\mathcal{C}$ does not depend on $p^{ij}$ (neither explicitly nor
through the $n^i$), the second line vanishes and the bracket is given
by the first line alone. The variations also include the dependence of
$\mathcal{C}$ on $\gamma_{ij}$ and $\pi^{ij}$ through $n^i$. However,
since $\partial\mathcal{C}/\partial n^k=0$ by \eqref{dR0/dn} and
\eqref{neqom}, the $n^i$ dependence can be ignored. Then, using
\eqref{bi-C}, one gets \cite{conf},
\begin{align} \label{CC-bracket}
\{ \mathcal{C}(x),\mathcal{C}(y) \} &= \{ R^{0(g)}(x),R^{0(g)}(y) \} + \{ R_i^{(g)}(x),R_j^{(g)}(y) \} {D^i}_k n^k(x) {D^j}_l n^l(y) \notag \\
&+ \{ R^{0(g)}(x),R_i^{(g)}(y) \} {D^i}_k n^k(y) - \{ R^{0(g)}(y),R_i^{(g)}(x) \} {D^i}_k n^k(x) \notag \\
&+ S^{mn}(x) \frac{\delta R_i^{(g)}(y)}{\delta \pi^{mn}(x)} {D^i}_k n^k(y) - S^{mn}(y) \frac{\delta R_i^{(g)}(x)}{\delta \pi^{mn}(y)} {D^i}_k n^k(x) \text{,}
\end{align}
where
\begin{equation} \label{Sdef}
S^{mn} = R_j^{(g)} \frac{\partial({D^j}_k n^k)}{\partial \gamma_{mn}} + 2m^4 \frac{\partial(\sqrt{\det \gamma} V)}{\partial \gamma_{mn}} \text{.}
\end{equation}
From equation \eqref{Ddef}, the following relations can be derived,
\begin{align} \label{dD/dg}
\frac{\partial}{\partial \gamma_{mn}} \Tr ( \sqrt{x} D) 
&= - \frac{1}{\sqrt{x}} \left( n^i \f_{ij} \frac{\partial ({D^j}_k n^k)}{\partial \gamma_{mn}} - \frac{1}{2} \f_{ij} {\left(D^{-1} \right)^j}_k \frac{\partial \gamma^{ki}}{\partial \gamma_{mn}} \right) \text { ,} \notag \\
\frac{\partial}{\partial \gamma_{mn}} \left( \Tr ( \sqrt{x} D)^2 \right) &= -2 \left( n^i \f_{ij} {D^j}_k \frac{\partial ({D^k}_l n^l)}{\partial \gamma_{mn}} - \frac{1}{2} \f_{ij} \frac{\partial \gamma^{ji}}{\partial \gamma_{mn}} \right) \text { ,} \\
\frac{\partial}{\partial \gamma_{mn}} \left( \Tr ( \sqrt{x} D)^3 \right) &= - 3 \sqrt{x} \left( n^i \f_{ij} {D^j}_k {D^k}_l \frac{\partial ({D^l}_r n^r)}{\partial \gamma_{mn}} - \frac{1}{2} \f_{ij} {D^j}_k \frac{\partial \gamma^{ki}}{\partial \gamma_{mn}} \right) \text { .} \notag
\end{align}
Using these, eliminating $R_j^{(g)}$ though \eqref{biC_i}, and
imposing equation \eqref{neqom}, $S^{mn}$ can be written as 
\begin{equation} \label{SV}
S^{mn} =  m^4 \sqrt{\det \gamma} \left(V \gamma^{mn} - \bar{V}^{mn} \right) \text{.}
\end{equation}
where,
\begin{align} \label{Vbardef}
\bar{V}^{mn} &= \gamma^{mi} \Bigg[ \beta_1 \frac{1}{\sqrt{x}} \f_{ik} {\left(D^{-1} \right)^k}_j + \beta_2 \left( \f_{ik} {\left(D^{-1} \right)^k}_j {D^l}_l - \f_{ij} \right) \notag \\
&\qquad\quad + \beta_3 \sqrt{x} \left( \f_{ik} {D^k}_j - \f_{ij} {D^k}_k + \frac{1}{2} \f_{ik} {\left(D^{-1} \right)^k}_j \left( {D^l}_l {D^h}_h - {D^l}_h {D^h}_l \right) \right) \Bigg] \gamma^{jn} \text{.}
\end{align}
Finally, using this, as well as the Poisson brackets
\eqref{intro_algebra},  one gets, 
\begin{equation} \label{CCbracket_final_form}
\{ \mathcal{C}(x),\mathcal{C}(y) \} = - \left[ \mathcal{C}(x) {D^i}_j n^j(x) \frac{\partial}{\partial x^i} \delta^3(x-y) - \mathcal{C}(y) {D^i}_j n^j(y) \frac{\partial}{\partial y^i} \delta^3(x-y) \right] \text{.}
\end{equation}
Hence, $\{ \mathcal{C}(x),\mathcal{C}(y) \} \approx 0$ which shows the
existence of the secondary constraint $\C\approx 0$ in massive
gravity, and potentially in bimetric theory \cite{conf}. $\C$ is given
by the remaining terms in \eqref{C2compbracket}.

\section{The secondary constraint in bimetric theory} \label{biseccon}

Here we compute the secondary constraint $\C$ in bimetric theory and 
show that its vanishing is independent of the lapse and shift
functions.  We also show that it reduces to the known expression for
massive gravity.

\subsection{Evaluation of the bimetric secondary constraint}

Using \eqref{CCbracket_final_form}, the expression
\eqref{C2compbracket} for $\C$ on the constraint surface becomes, 
\begin{align} \label{C2bracket}
\mathcal{C}_{(2)}(z) &\approx - \int \mathrm{d}^3y \,
\{\mathcal{C}(z), L^i(y) \tilde{R}_i(y) + L(y) \tilde{R}^0(y) \} 
\\
 &=- \int \mathrm{d}^3y \left( \{\mathcal{C}(z),L^i(y) \tilde{R}_i(y) +
L(y) \tilde{R}^0(y) \}_g + \{\mathcal{C}(z),L^i(y) \tilde{R}_i(y) +
L(y) \tilde{R}^0(y) \}_f   \right) \text{.} \notag
\end{align}
In the second line, the brackets $\{\}_g$ are evaluated with respect
to $(\gamma_{ij},\pi^{ij})$, and $\{\}_f$ are evaluated with respect
to $(\f_{ij},p^{ij})$. Note that, as before, the dependence of
$\mathcal{C}$ and $\tilde{R}^0$ on $\gamma_{ij}$, $\pi^{ij}$ and
$\f_{ij}$ through $n^i$ can be ignored, due to \eqref{dR0/dn}. The
brackets are therefore evaluated at fixed $n^i$. By inspecting
\eqref{R0tilde_def} and \eqref{Ritilde_def}, we observe that the $\{
\}_g$ bracket coincides with a Poisson bracket computed in \cite{conf} 
for massive gravity, since the new terms in $L^i \tilde{R}_i + L
\tilde{R}^0$ that contain $R^{0 (f)}$ and $R_i^{(f)}$ do not
contribute to the bracket. This contribution is given by,
\begin{align} \label{CLRg_bracket}
\int \mathrm{d}^3y & \{\mathcal{C}(z),L^i(y) \tilde{R}_i(y) + L(y) \tilde{R}^0(y) \}_g = - \Bigg( \frac{m^4}{M_g^2} L \left(\gamma_{mn} {\pi^k}_k - 2 \pi_{mn} \right) U^{mn} \notag \\
& + 2m^4 \sqrt{\det \gamma} \gamma_{ni} {D^i}_k n^k \nabla_m \left( L U^{mn} \right)  + \left( R_j^{(g)} {D^i}_k n^k - 2m^4 \sqrt{\det \gamma} \gamma_{jk} \bar{V}^{ki} \right) \nabla_i (Ln^j+L^j) \notag \\
& + \sqrt{\det \gamma} \left[ \nabla_i \left( \frac{R^{0(g)}}{\sqrt{\det \gamma}} \right) + \nabla_i \left( \frac{R_j^{(g)}}{\sqrt{\det \gamma}} \right) {D^j}_k n^k \right] (Ln^i+L^i) \Bigg) \text{,}
\end{align}
where
\begin{equation} \label{Umn-def}
U^{mn} \equiv \frac{2}{\sqrt{\det \gamma}} \frac{\delta(\sqrt{\det
    \gamma} U)}{\delta \gamma_{mn}} 
 = U \gamma^{mn} + 2 \frac{\partial U}{\partial \gamma_{mn}}\text{.}
\end{equation}
As for the $\{\}_f$ bracket, since $\mathcal{C}$ is independent of
$p^{ij}$, the only nonzero contributions will come from the $p^{ij}$ 
dependent terms in $\tilde{R}^0$ and $\tilde{R}_i$, combined with
$\f_{ij}$ dependent terms in $\mathcal{C}$, 
\begin{equation} \label{CHfbracket}
\{\mathcal{C}(z),L^i(y) \tilde{R}_i(y) + L(y) \tilde{R}^0(y) \}_f   =  Z^{mn}(z) \left(  L^k(y) \frac{\delta R_k^{(f)}(y)}{\delta p^{mn}(z)} + L(y) \frac{\delta R^{0 (f)}(y)}{\delta p^{mn}(z)} \right) \text{,}
\end{equation}
where $Z^{mn}$ is given by,
\begin{equation} \label{Zdef}
Z^{mn} = \frac{\partial \mathcal{C}}{\partial \f_{mn}} = R_j^{(g)} \frac{\partial({D^j}_k n^k)}{\partial \f_{mn}} + 2m^4 \sqrt{\det \gamma} \frac{\partial V}{\partial \f_{mn}} \text{.}
\end{equation}
To compute $Z^{mn}$, we need the following expressions that can be
derived from \eqref{Ddef},  
\begin{align} \label{dD1}
\frac{\partial}{\partial \f_{mn}} \Tr ( \sqrt{x} D) 
&= -\frac{1}{\sqrt{x}} \left[ n^i \f_{ij} \frac{\partial \left({D^j}_k
    n^k \right)}{\partial \f_{mn}} - \frac{1}{2} Q_1^{mn}
  \right]\text{,}  
\\ \label{dD2}
\frac{\partial}{\partial \f_{mn}} \Tr (xD^2) 
&= -2 \left[ {D^i}_l n^l \f_{ij} \frac{\partial \left( {D^j}_k n^k
    \right)}{\partial \f_{mn}} - \frac{1}{2} Q_2^{mn} \right] \text{,} 
\\ \label{dD3}
\frac{\partial}{\partial \f_{mn}} \Tr (x^{3/2} D^3)
&= -3 \sqrt{x}\left[ {D^i}_l {D^l}_r n^r \f_{ij} \frac{\partial 
\left( {D^j}_k n^k \right) }{\partial \f_{mn}} - \frac{1}{2} Q_3^{mn}
  \right] \text{.} 
\end{align}
where $Q_1^{mn}$, $Q_2^{mn}$ and $Q_3^{mn}$ are defined as,
\begin{align} \label{Q1def}
Q_1^{mn} &= {( D^{-1} )^{m}}_i \gamma^{ni} - n^i n^{m} {D^{n}}_i \text{,}
\\
\label{Q2def}
Q_2^{mn} &= \gamma^{mn} - {D^m}_i n^i {D^n}_j n^j \text{,}
\\
\label{Q3def}
Q_3^{mn} &= {D^m}_i \gamma^{in} - {D^m}_i {D^i}_j n^j {D^n}_k n^k \text{,}
\end{align}
and can be shown to be symmetric using \eqref{Ddef}.  Differentiating
equation \eqref{Vdef} with respect to $f_{mn}$, using
\eqref{dD1}-\eqref{dD3} and performing some algebraic manipulations
yields,
\begin{align} \label{dV-W}
\frac{ \partial V}{\partial \f_{mn}} = - \frac{n^l \f_{lk}}{\sqrt{x}} \bigg[ & 
\beta_1 {\delta^k}_i + \beta_2 \sqrt{x} \left( {\delta^k}_i {D^r}_r - {D^k}_i \right) \notag \\
&+ \beta_3 x \left( {\delta^k}_i  \, e_2(D)  + {D^k}_r {D^r}_i - {D^k}_i {D^r}_r \right) \bigg] \frac{\partial \left( {D^i}_j n^j \right)}{\partial \f_{mn}} + \frac{1}{2} \bar{W}^{mn} \text{,}
\end{align}
where
\begin{align} \label{WQ}
\bar{W}^{mn} = \frac{\beta_1}{\sqrt{x}} Q_1^{mn} 
+ \beta_2 \left( {D^i}_i Q_1^{mn} - Q_2^{mn} \right) 
+ \beta_3 \sqrt{x} \left( e_2(D) Q_1^{mn} - {D^i}_i Q_2^{mn} +
Q_3^{mn} \right) \text{.} 
\end{align}
On inserting \eqref{dV-W} into \eqref{Zdef} and imposing
\eqref{neqom}, $Z^{mn}$ becomes,
\begin{equation} \label{ZW}
Z^{mn} = m^4 \sqrt{\det \gamma} \bar{W}^{mn} \text{.}
\end{equation}
Now, to integrate \eqref{CHfbracket}, we use
\begin{equation} 
\frac{\delta R^{0 (f)}(y)}{\delta p^{mn}(z)} = \frac{1}{M_f^2
  \sqrt{\det \f}} \left( \f_{mn}(y) {p^k}_k(y) - 2 p_{mn}(y) \right)
\delta^3(z-y) \text{,} 
\end{equation}
\begin{equation}
\int \mathrm{d}^3y L^i(y) \frac{\delta R_i^{(f)}(y)}{\delta p^{mn}(z)} = - \left[  (\bar{\nabla}_z)_m L_n(z) +  (\bar{\nabla}_z)_n L_m(z) \right]  \text { ,}
\end{equation}
which are similar to the corresponding expressions for $R^{0 (g)}$ and
$R_i^{(g)}$ in \cite{conf}. Performing the integration, and using
equation \eqref{ZW}, yields 
\begin{align}
\label{f-bracket}
- &\int \mathrm{d}^3y \{\mathcal{C}(z),L^i(y) \tilde{R}_i(y) + L(y) \tilde{R}^0(y) \}_f  \notag \\ 
&\qquad\qquad =  m^4 \sqrt{\det \gamma} \bar{W}^{mn}  \left( \frac{2L}{M_f^2
  \sqrt{\det \f}}  \left( p_{mn} - \frac{1}{2} \f_{mn} {p^k}_k \right) +  \bar{\nabla}_m L_n + \bar{\nabla}_n L_m  \right) \text{.}
\end{align}
Combining this with \eqref{CLRg_bracket} gives a preliminary
expression for the secondary constraint in bimetric theory which
seemingly depends on $L$, $L_i$ and their spatial derivatives,
\begin{align} \label{bimetricC2}
\mathcal{C}_{(2)}  \approx &\frac{m^4}{M_g^2} L \left(\gamma_{mn} {\pi^k}_k - 2 \pi_{mn} \right) U^{mn} + 2m^4 \sqrt{\det \gamma} \gamma_{ni} {D^i}_k n^k \nabla_m \left( L U^{mn} \right) \notag \\
 & + \left( R_j^{(g)} {D^i}_k n^k - 2m^4 \sqrt{\det \gamma} \gamma_{jk} \bar{V}^{ki} \right) \nabla_i (Ln^j+L^j) \notag \\
 & + \sqrt{\det \gamma} \left[ \nabla_i \left( \frac{R^{0 (g)}}{\sqrt{\det \gamma}} \right) + \nabla_i \left( \frac{R_j^{(g)}}{\sqrt{\det \gamma}} \right) {D^j}_k n^k \right] (Ln^i+L^i) \notag \\
 & - \frac{m^4}{M_f^2} L \frac{\sqrt{\det \gamma}}{\sqrt{\det \f}} \left( \f_{mn} {p^k}_k - 2 p_{mn} \right) \bar{W}^{mn} + 2m^4 \sqrt{\det \gamma} \f_{jk} \bar{W}^{ki} \bar{\nabla}_i L^j \text{.}
\end{align}

One can verify that $\C$ is consistent with the corresponding
expression in massive gravity. For this, note that the last line of
\eqref{bimetricC2} coming from \eqref{f-bracket} is related to the
$\f_{mn}$ equation of motion,
\begin{equation} \label{3feom}
\frac{\partial}{\partial t} \f_{mn} = \frac{2L}{M_f^2 \sqrt{\det \f}} \left( p_{mn} - \frac{1}{2} \f_{mn} {p^k}_k \right) +  \bar{\nabla}_m L_n + \bar{\nabla}_n L_m \text{.}
\end{equation}
Hence, using \eqref{Zdef} and \eqref{ZW}, we have,
\begin{equation} \label{dC/df*df/dt}
-\int \mathrm{d}^3y \{\mathcal{C}(z),L^i(y) \tilde{R}_i(y) + L(y) \tilde{R}^0(y) \}_f = \frac{\partial \mathcal{C}}{\partial \f_{mn}} \frac{\partial}{\partial t} \f_{mn} \text{.}
\end{equation}
In massive gravity where $\f_{mn}$ is nondynamical, the right hand
side corresponds to the explicit time dependence of $\mathcal{C}$
through the $f_{\mu\nu}$ background. More precisely, consider,
\begin{equation} \label{C2_chainrule}
\mathcal{C}_{(2)} = \frac{d}{dt} \mathcal{C} = \frac{\partial \mathcal{C}}{\partial \gamma_{ij}} \frac{\partial \gamma_{ij}}{\partial t} + \frac{\partial \mathcal{C}}{\partial \pi^{ij}} \frac{\partial \pi^{ij}}{\partial t} + \frac{\partial \mathcal{C}}{\partial \f_{ij}} \frac{\partial \f_{ij}}{\partial t} + \frac{\partial \mathcal{C}}{\partial p^{ij}} \frac{\partial p^{ij}}{\partial t} \text{.}
\end{equation}
The last term vanishes since $\mathcal{C}$ does not depend on
$p^{ij}$, and the third term is the one in \eqref{dC/df*df/dt}. The 
first two terms produce the Poisson bracket with respect to
$(\gamma_{ij},\pi^{ij})$ and are common with massive gravity.
Then, from \eqref{bimetricC2}, the expression for $\C$ in massive
gravity becomes,  
\begin{align} \label{massiveC2}
\mathcal{C}_{(2)}^\text{MG}  \approx &\frac{\partial \mathcal{C}}{\partial \f_{ij}} \frac{\partial \f_{ij}}{\partial t} + \frac{m^4}{M_g^2} L \left(\gamma_{mn} {\pi^k}_k - 2 \pi_{mn} \right) U^{mn} + 2m^4 \sqrt{\det \gamma} \gamma_{ni} {D^i}_k n^k \nabla_m \left( L U^{mn} \right) \notag \\
& + \left( R_j^{(g)} {D^i}_k n^k - 2m^4 \sqrt{\det \gamma} \gamma_{jk} \bar{V}^{ki} \right) \nabla_i (Ln^j+L^j) \notag \\
& + \sqrt{\det \gamma} \left[ \nabla_i \left( \frac{R^{0(g)}}{\sqrt{\det \gamma}} \right) + \nabla_i \left( \frac{R_j^{(g)}}{\sqrt{\det \gamma}} \right) {D^j}_k n^k \right] (Ln^i+L^i) \text{,}
\end{align}
which is consistent with the expression in \cite{conf}.\footnote{In
  \cite{conf}, the $L$ in the third term of \eqref{massiveC2} is
  outside the covariant derivative which is a misprint corrected
  here. Also the first term is absent for backgrounds with
  $f_{\mu\nu}=\eta_{\mu\nu}$.}

\subsection{The $L^i$ independence of $\mathcal{C}_{(2)}$} \label{Li-dep}

Note that all $L^i$ dependent terms of $\C$ arise from the bracket
$\{\mathcal{C}(z), L^i(y) \tilde{R}_i(y)\}$ in equation \eqref{C2bracket}. In
this section, we show that, on the constraint surface, all such terms
vanish in bimetric theory (though not in massive gravity). Denoting
the $L^i$ dependent part of $\mathcal{C}_{(2)}$ in \eqref{bimetricC2}
by $\mathcal{C}_S$, one gets,
\begin{align} \label{CS_def}
\mathcal{C}_S & \approx 
\sqrt{\det \gamma} \left[ \nabla_i \Big( \frac{R^{0 (g)}}{\sqrt{\det
      \gamma}} \Big) + \nabla_i \Big( \frac{R_j^{(g)}}{\sqrt{\det
      \gamma}} \Big) {D^j}_k n^k \right] L^i 
\notag \\
&\qquad +R_j^{(g)} {D^i}_k n^k \nabla_i L^j 
- 2m^4 \sqrt{\det \gamma} \left(\gamma_{jk} \bar{V}^{ki} \nabla_i L^j 
-\f_{jk} \bar{W}^{ki} \bar{\nabla}_i L^j \right) \text{.}
\end{align}
By adding and subtracting terms, the first line can be completed to
include ${\cal C}$ \eqref{bi-C}, and we get,
\begin{align} \label{CS_C}
\mathcal{C}_S & \approx L^i \partial_i {\cal C} - \Delta \,.
\end{align}
To write the first term, we have used the fact that $\cal C$ is a
scalar density so that ${\cal C}/{\sqrt{\det\gamma}}$ is a
scalar. Hence, one has $\sqrt{\det\gamma}\nabla_i ({\cal
  C}/{\sqrt{\det\gamma}})\approx \partial_i{\cal C}$, on the
constraint surface. Since ${\cal C}\approx 0$ at all points on a
spacelike hypersurface, it follows that $\partial_i{\cal C}\approx
0$. In the second term, the quantity $\Delta$ stands for,
\begin{align} \label{Delta}
\Delta =&
R_j^{(g)}\left(\nabla_i({D^j}_k n^k)L^i - {D^i}_k n^k\nabla_i L^j\right) 
\notag\\ 
&+2m^4\sqrt{\det\gamma}
\left( \partial_i V L^i+ \gamma_{jk} \bar{V}^{ki} \nabla_i L^j 
 -\f_{jk} \bar{W}^{ki} \bar{\nabla}_i L^j \right)  \text{.}
\end{align}
Since $\int \diff^3y \{\mathcal{C}(z), L^i(y) \tilde{R}_i(y)\}=-{\cal C}_S(z)$, the quantity
$\Delta$ appears in the computation of this bracket in Appendix
\ref{Cbracketsection}, more precisely, in equation
\eqref{FCGbracket-C}. A lengthy calculation, which is relegated to the
appendix, shows that $\Delta=0$ and therefore, ${\cal C}_S\approx 0$.
The dependence on $L^i$ therefore vanishes entirely and all such terms
should be dropped from the expression for $\C$ in \eqref{bimetricC2}.

\subsection{The $\partial_i L$ independence of $\C$} \label{delL-dep}

In this section, it is shown that all terms in $\C$ involving
derivatives of $L$ vanish on the constraint surface. Denote the
$\partial_i L$ dependent part of $\C$ by $\mathcal{C}_D$. From
equation \eqref{bimetricC2} it follows that
\begin{align} \label{CD}
\mathcal{C}_D =& \left( R_j^{(g)} n^j 
+ 2m^4 \sqrt{\det \gamma} U  \right)   {D^i}_k n^k \partial_i L 
\\
& + 4m^4 \sqrt{\det \gamma} \gamma_{ni} {D^i}_k n^k
\frac{\partial U}{\partial \gamma_{mn}}\partial_m L 
-2m^4\sqrt{\det\gamma}\gamma_{in}\bar{V}^{nm} n^i\partial_m L 
\text{,}  
\end{align}
where we have used the expression \eqref{Umn-def} for $U^{mn}$. The
first two terms are simplified by using \eqref{Udef} and \eqref{xdef}
to rewrite $U$, and then imposing \eqref{neqom} to get,
\begin{align} \label{CDstep1}
\left( R_j^{(g)}n^j+2m^4\sqrt{\det\gamma} U\right)
= 2m^4\sqrt{\det\gamma}\left[ \frac{\beta_1}{\sqrt{x}}+\beta_2 {D^l}_l
+ \beta_3 \sqrt{x}\, e_2(D)\right]\text{.}
\end{align}
Note that $U$ given in \eqref{Udef} has the form $U=\beta_1 [U_1] +
\beta_2 [U_2] + \beta_3 [U_3]$. To simplify the third term in
$\mathcal{C}_D$, we need to compute $\frac{\partial
  U}{\partial\gamma_{mn}}$ (at fixed $n_i$). The $\beta_1$-term
vanishes trivially,  
\begin{equation}\label{U1}
\beta_1 \frac{\partial [U_1]}{\partial \gamma_{mn}} =
\beta_1 \frac{\partial \sqrt{x}}{\partial \gamma_{mn}} = 0 \text{.}
\end{equation}
The $\beta_2$-term is also easily evaluated using the first equation
in \eqref{dD/dg},
\begin{align} \label{U2}
\beta_2   \frac{\partial [U_2]}{\partial \gamma_{mn}} 
= \frac{\beta_2}{2} \f_{ij} { \left(D^{-1} \right)^j}_k \frac{
  \partial \gamma^{ki}}{\partial \gamma_{mn}}\text{.} 
\end{align}
Similarly, using \eqref{dD/dg}, the $\beta_3$-term becomes, 
\begin{align}\label{U3}
\beta_3 \frac{\partial [U_3]}{\partial \gamma_{mn}} 
=\frac{\beta_3}{2\sqrt{x}}\Bigg[x{D^l}_l {\left(D^{-1} \right)^j}_k 
-{x} \delta^j_k 
+n^l \f_{lq} {D^q}_r n^r {\left(D^{-1}\right)^j}_k
-n^l \f_{lk} n^j \Bigg]
\f_{ji} \frac{\partial \gamma^{ki}}{\partial \gamma_{mn}} 
\text{,}
\end{align}
where we have also used the relation,\footnote{Equation \eqref{simp_g}
  is analogous to equation (4.6) in \cite{fra}, and is derived in a
  similar way.}  
\begin{align} \label{simp_g}
& n^l \f_{li} \left(x {D^i}_j + {D^i}_r n^r n^q \f_{qj} \right)
  \frac{\partial\left({D^j}_k n^k \right)}{\partial\gamma_{mn}}  =
  \frac{1}{2} n^i \f_{ij} n^l \f_{lk} \frac{\partial\gamma^{jk}}
       {\partial \gamma_{mn}}   \text{.}  
\end{align}
Now, substituting \eqref{CDstep1}, \eqref{U1}, \eqref{U2} and \eqref{U3} in
the expression for $\mathcal{C}_D$, one can see that it vanishes on
using the identity, 
\begin{equation} \label{D3eq}
{D^i}_k n^k - n^j \f_{jm} {\left(D^{-1} \right)^m}_n \gamma^{ni} = 0 \text{,}
\end{equation}
which follows from equations \eqref{Ddef} and \eqref{xdef} that define
the matrix $D$. For example, the $\beta_1$-term of $\mathcal{C}_D$ is
given by,  
\begin{align}
\mathcal{C}_D^{\beta_1} = 2 m^4 \sqrt{\det \gamma} \frac{\beta_1}{\sqrt{x}} \left[ {D^i}_k n^k - n^j \f_{jm} {\left(D^{-1} \right)^m}_n \gamma^{ni} \right] \partial_i L = 0 \text{,}
\end{align}
Similarly one can show that $\mathcal{C}_D^{\beta_2}$ and
$\mathcal{C}_D^{\beta_3}$ also vanish on repeated application of
\eqref{D3eq}, hence, $\mathcal{C}_D = 0$. This means that $\C$ 
is independent of $\partial_i L$ and can be expressed as, 
\begin{align} \label{bimetricC2_final_form}
\mathcal{C}_{(2)} & \approx L \Bigg( \frac{m^4}{M_g^2} \left(\gamma_{mn} {\pi^k}_k - 2 \pi_{mn} \right) U^{mn} + 2m^4 \sqrt{\det \gamma} \gamma_{ni} {D^i}_k n^k \nabla_m   U^{mn}  \notag \\
& + \left( R_j^{(g)} {D^i}_k n^k - 2m^4 \sqrt{\det \gamma} \gamma_{jk} \bar{V}^{ki} \right) \nabla_i n^j \notag \\
& + \sqrt{\det \gamma} \left[ \nabla_i \left( \frac{R^{0 (g)}}{\sqrt{\det \gamma}} \right) + \nabla_i \left( \frac{R_j^{(g)}}{\sqrt{\det \gamma}} \right) {D^j}_k n^k \right] n^i \notag \\
& - \frac{m^4}{M_f^2} \frac{\sqrt{\det \gamma}}{\sqrt{\det \f}} \left( \f_{mn} {p^k}_k - 2 p_{mn} \right) \bar{W}^{mn} \Bigg) \text{.}
\end{align}
Since $L$ is an overall factor, imposing this secondary constraint is
equivalent to $\C/L \approx 0$. This can be solved for the canonical
momentum of the ghost mode.

One can check that there are no additional secondary constraints. 
From \eqref{R0Ri-tilde} and \eqref{R0R0-tilde} below, we see that the 
brackets $\{ \tilde{R}^0(x),\tilde{R}_i(y) \}$ and
$\{\tilde{R}^0(x),\tilde{R}^0(y) \}$ vanish on the constraint
surface. Thus, the condition $\dot{\tilde{R}}^0 \approx 0$, leads 
to an equation that is identical with \eqref{bimetricC2_final_form},
(except that $L$ is replaced by $N$) and imposes the same constraint.
This, combined with the fact that $\dot{\tilde{R}}_i \approx 0$ is
automatically satisfied, shows that there are no further constraints.

\section{The algebra of general covariance and the HKT
  metric} \label{algebra-chapter}   

\subsection{The algebra of general covariance} \label{algebra}

In general relativity, the Poisson brackets of the first class
constraints, $R_0\approx 0$, $R_i\approx 0$, are \cite{DeWitt:1967yk},
\begin{align} \label{GRalgebra}
\{R^0(x) , R^0(y) \} &= R^i(y) \frac{\partial}{\partial y^i} \delta^3(x-y)
- R^i(x) \frac{\partial}{\partial x^i} \delta^3(x-y) \text{,} \notag \\
\{R^0(x) , R_i(y) \} &= - R^0(y) \frac{\partial}{\partial x^i} 
\delta^3(x-y) \text{,} \\
\{R_i(x) , R_j(y) \} &= R_i(y) \frac{\partial}{\partial y^j} \delta^3(x-y)
- R_j(x) \frac{\partial}{\partial x^i} \delta^3(x-y)  
\text{.}\notag
\end{align}
This is the algebra of spacetime diffeomorphisms and is not unique to
general relativity. Any covariant field theory will have first class
constraints satisfying the algebra \eqref{GRalgebra} after second
class constraints have been imposed \cite{Teitelboim}. This is a
necessary and sufficient condition for the constraints to generate
spacetime diffeomorphisms.

Since the algebra \eqref{GRalgebra} contains both $R_i$ and $R^i$, it
implicitly contains a metric to manipulate the index. In general
relativity and theories minimally coupled to it, this is the spatial
metric $\gamma_{ij}\equiv g_{ij}$. Furthermore, in such theories, the
diffeomorphism constraints appear in the Hamiltonian with Lagrange
multipliers, as 
\begin{equation} \label{firstclass_Hamiltonian}
\mathcal{H} = -N R^0- N^i R_i 
\text{,}
\end{equation}
provided second class constraints and their stability conditions,
if any, are already imposed. Here, $N$ and $N^i$ are the lapse and
shift functions that complete $g_{ij}$ to $g_{\mu\nu}$.

This observations can help identify a spacetime metric in a
diffeomorphism invariant theory. The algebra \eqref{GRalgebra} also
holds in generally covariant theories where a spacetime metric is not
{\it a priori} specified. Then, as conjectured by Hojman, Kucha\v{r}
and Teitelboim \cite{hkt}, the tensor that relates $R_i$ and $R^i$ can
be identified as the metric of spacelike hypersurfaces. An explicit
example of this, for Ashtekar's canonical formulation of gravity, is
given in \cite{Peldan}. Here we are interested in the implications of
the conjecture for bimetric theory which does not {\it a priori} have
a unique metric candidate. We will obtain the algebra of first class
constraints and show that it leads to different HKT metrics depending
on how the constraints are identified. Hence in these type of models
the HKT conjecture does not lead to a unique metric. 

\subsection{First class constraints in bimetric theory}

Bimetric theory has six independent constraints, $\tilde{R}^0$,
$\tilde{R}_i$, $\mathcal{C}$ and $\C$, given by equations
\eqref{R0tilde_def} - \eqref{bi-C} and
\eqref{bimetricC2_final_form}. Four linear combinations of these
should be first class and satisfy the algebra \eqref{GRalgebra}. We
first show that $\tilde{R}_i$ are first class. The Poisson brackets of
the first five constraints are 
\begin{align}
&\{  \tilde{R}_i(x), \tilde{R}_j(y) \} = - \left[ \tilde{R}_j(x) \frac{\partial}{\partial x^i} \delta^3(x-y) - \tilde{R}_i(y) \frac{\partial}{\partial y^j} \delta^3(x-y) \right] \text{,} \label{RiRj-tilde} \\
&\{ \tilde{R}^0(x),\tilde{R}_i(y) \} = - \tilde{R}^0(y) \frac{\partial}{\partial x^i} \delta^3(x-y) \text{,} \label{R0Ri-tilde} \\
&\{ \tilde{R}^0(x),\tilde{R}^0(y) \} = - \left[ \f^{ij}(x) \tilde{R}_j(x) \frac{\partial}{\partial x^i} \delta^3(x-y) - \f^{ij}(y) \tilde{R}_j(y) \frac{\partial}{\partial y^i} \delta^3(x-y) \right] \text{,} \label{R0R0-tilde} \\
&\{ \mathcal{C}(x),\tilde{R}_i(y) \} = - \mathcal{C}(y) \frac{\partial}{\partial x^i} \delta^3(x-y) \text{,} \label{CRibracket} \\
&\{ \mathcal{C}(x),\mathcal{C}(y) \} = - \left[ \mathcal{C}(x) {D^i}_j n^j(x) \frac{\partial}{\partial x^i} \delta^3(x-y) - \mathcal{C}(y) {D^i}_j n^j(y) \frac{\partial}{\partial y^i} \delta^3(x-y) \right] \text{,} \label{CCalgebra} \\
&\{ \mathcal{C}(x), \tilde{R}^0(y) \} = - \frac{\C(x)}{L(x)} \delta^3(x-y) \text{.} \label{CR0_text}
\end{align}
Equation \eqref{RiRj-tilde} is easily derived from
\eqref{Ritilde_def}, and the fact that $R_i^{(g)}$ and $R_i^{(f)}$
each satisfy the spatial part of \eqref{GRalgebra}. The bracket
\eqref{CCalgebra} is given in \eqref{CCbracket_final_form} and the
remaining ones are derived in appendix \ref{appendix_brackets}. All of
these brackets vanish on the constraint surface. 

Now consider the brackets involving $\C$, starting with $\{
\C(x),\tilde{R}_i(y) \}$. This vanishes on the constraint 
surface as seen from the definition of $\C$ in \eqref{C2def} and
the Jacobi identity,
\begin{align}
\{ \C(x),\tilde{R}_i(y) \} = \{ \{ \mathcal{C}(x),H\},\tilde{R}_i(y)\}
=\{ \mathcal{C}(x),\{ H ,\tilde{R}_i(y) \} \} -\{ H,\{\mathcal{C}(x),
\tilde{R}_i(y) \} \} \text{.} 
\end{align}
Indeed, using \eqref{CRibracket}, the last bracket becomes,
\begin{equation}
\{ H , \{ \mathcal{C}(x),\tilde{R}_i(y) \} \}=\C(y)\frac{\partial}{\partial x^i}
\delta^3(x-y) \text{,}
\end{equation}
which is weakly zero. In addition, using \eqref{bi-Hamiltonian}, as
well as \eqref{RiRj-tilde}, \eqref{R0Ri-tilde} and \eqref{CRibracket}
it follows that 
\begin{align}
\{ H ,\tilde{R}_i(y) \} = - \left[ \frac{\partial N(y)}{\partial y^i} \mathcal{C}(y) + \frac{\partial L(y)}{\partial y^i} \tilde{R}^0(y) + \frac{\partial}{\partial y^j} \left( L^j(y) \tilde{R}_i(y) \right) + \frac{\partial L^j(y)}{\partial y^i} \tilde{R}_j(y) \right] \text{.}
\end{align}
Now all brackets in $\{ \mathcal{C}(x),\{ H ,\tilde{R}_i(y) \} \}$ are 
known and result in terms proportional to a constraint or its
spatial derivative. Hence, this expression, and therefore
$\{\C(x),\tilde{R}_i(y) \}$, is weakly zero, showing that the
Poisson brackets of $\tilde{R}_i$ with all constraints vanish weakly.
Therefore, $\tilde{R}_i$ are first class constraints generating the
algebra of spatial diffeomorphisms in \eqref{GRalgebra}.

To find the remaining first class constraint, we turn our attention to
the brackets $\{ \mathcal{C}(x),\C(y) \}$ and $\{ \tilde{R}^0(x),\C(y)
\}$, neither of which is weakly zero. This can be seen by noting that
they appear in the expression for $\dot{\mathcal{C}}_{(2)}$, which, on
using \eqref{bi-Hamiltonian} and $\{\C(x),\tilde{R}_i(y) \}\approx 0$,
becomes,
\begin{equation} \label{C2dot}
\dot{\mathcal{C}}_{(2)}(x) \approx  -\int \mathrm{d}y \left[ L(y) \{\mathcal{C}_{(2)}(x),\tilde{R}^0(y) \} + N(y) \{\mathcal{C}_{(2)}(x),\mathcal{C}(y) \} \right] \text{.}
\end{equation}
This is computed at the linear level in appendix \ref{det_Lagrange}
and the result, in equation \eqref{linearC2dot}, shows that the two
brackets are not weakly zero. Hence, none of the constraints
$\tilde{R}^0$, $\mathcal{C}$ or $\C$ is first class.

However, a first class combination of them could exist
\cite{Dirac1964}. Note that the brackets $\{ \mathcal{C}(x),\C(y) \}$
and $\{ \tilde{R}^0(x),\C(y) \}$ contain no derivatives of delta
functions, as can be shown using the Jacobi identity, similar to the
analysis in \cite{krasnov}. Then they have the form,
\begin{equation} \label{CC2bracket}
\{ \mathcal{C}(x),\C(y) \} = L(x) D(x) \delta^3(x-y) \text{,}
\end{equation}
\begin{equation} \label{R0C2bracket}
\{ \tilde{R}^0(x),\C(y) \} = L(x) E(x) \delta^3(x-y) \text{,}
\end{equation}
where $D$ and $E$ are nonzero functions of the phase space
variables. Now define the constraint $\hat{R}^0$ as the linear
combination, 
\begin{equation}
\hat{R}^0 \equiv W \mathcal{C} + \tilde{R}^0 \text{,} \qquad W \equiv -\frac{E}{D} \text{.}
\end{equation}
The Poisson bracket of $\hat{R}^0$ and $\C$ is now weakly zero, 
\begin{align}
\{ \hat{R}^0(x),\C(y) \} &= \{ W(x) , \C(y) \} \mathcal{C}(x) + W(x) \{ \mathcal{C}(x) , \C(y) \} + \{ \tilde{R}^0(x) , \C(y) \} \notag \\
& \approx -\frac{E(x)}{D(x)} L(x) D(x) \delta^3(x-y) + L(x) E(x) \delta^3(x-y) = 0 \text{.}
\end{align}
Moreover, from \eqref{R0Ri-tilde} - \eqref{CR0_text}, it follows that
the brackets of $\hat{R}^0$ with other constraints also vanish on the
constraint surface. Hence, $\hat{R}^0$ is a first class
constraint. Note that we will find the same constraint by solving the
stability condition $\dot{\mathcal{C}}_{(2)} \approx 0$ for $N$ and
replacing it in the Hamiltonian. Indeed, using \eqref{CC2bracket} and
\eqref{R0C2bracket} in \eqref{C2dot} and setting
$\dot{\mathcal{C}}_{(2)} \approx 0$ gives $N = WL$. This, when
inserted into the Hamiltonian \eqref{bi-Hamiltonian}, reduces it to
\begin{equation} \label{fHamiltonian}
\mathcal{H} = - L \hat{R}^0 - L^i \tilde{R}_i \text{,}
\end{equation}
which is the desired form \eqref{firstclass_Hamiltonian}. Furthermore,
using \eqref{R0Ri-tilde} - \eqref{CR0_text}, it follows that  
\begin{equation}
\{ \hat{R}^0(x),\tilde{R}_i(y) \} \approx - \hat{R}^0(y) \frac{\partial}{\partial x^i} \delta^3(x-y) \text{.}
\end{equation}
\begin{equation} \label{R0R0-hat}
\{ \hat{R}^0(x),\hat{R}^0(y) \} \approx - \left[ \f^{ij}(x) \tilde{R}_j(x) \frac{\partial}{\partial x^i} \delta^3(x-y) - \f^{ij}(y) \tilde{R}_j(y) \frac{\partial}{\partial y^i} \delta^3(x-y) \right] \text{.}
\end{equation}
The weak equalities $\approx$ now hold on the surface of the second
class constraints $\mathcal{C}$ and $\C$. $\hat{R}^0$ generates time
diffeomorphisms as in \eqref{GRalgebra}. $\hat{R}^0$ and $\tilde{R}_i$
together generate the diffeomorphism algebra. 

\subsection{The HKT metrics}

It is obvious that on the right hand side of \eqref{R0R0-hat}, the
index of $\tilde{R}_i$ is raised using $\f^{ij}$, hence, by the HKT
conjecture \cite{hkt} described above, $f_{ij}\equiv\f_{ij}$ will be
a HKT spatial metric in bimetric theory. Furthermore, the Lagrange
multipliers $L$ and $L^i$ in the Hamiltonian \eqref{fHamiltonian} will
be the corresponding lapse and shift functions, leading to a spacetime
metric $f_{\mu\nu}$.

Singling out $f_{\mu\nu}$ in this way might seem odd considering that
both metrics appear in the bimetric action on similar footing. But
this is a consequence of the choice of variables that was made to
arrive at the constraints. Obviously, it is possible to repeat all
manipulations with $g_{\mu \nu}$ and $f_{\mu \nu}$ interchanged, as
commented in footnote \ref{fn-symmetry}. This will result in a
Lagrangian linear in $L$, $N$ and $N^i$, and finally $\gamma_{ij}$
will appear as the HKT metric in the covariance algebra.

In fact, it is not even necessary to repeat all steps with the $g$ and
$f$ metrics interchanged. We can start with the $n^i$ still given by
\eqref{ndef} but use them to eliminate the $L^i$ (instead of the
$N^i$). The resulting Lagrangian has the form,
\begin{equation}
\mathcal{L} = \pi^{ij} \partial_t \gamma_{ij} + p^{ij} \partial_t \f_{ij} + N^i \tilde{R}_i + L \mathcal{C}_L + N \mathcal{C}_N \text{,}
\end{equation}
where $\mathcal{C}_N$ and $\mathcal{C}_L$ differ from $\mathcal{C}$
and $\tilde{R}^0$. Then, proceeding as in the previous section, the
stability conditions of $\mathcal{C}_N$ and $\mathcal{C}_L$ provide a 
sixth constraint, whose stability condition in turn can be solved for
$L$. Using this solution gives a Hamiltonian $\mathcal{H} = -N
\tilde{\tilde R}^0- N^i \tilde{R}_i$, where $\tilde{\tilde R}^0$ and
$\tilde{R}_i$ are first class constraints satisfying the spatial 
diffeomorphism algebra and,  
\begin{equation}
\{\tilde{\tilde R}^0(x),\tilde{\tilde  R}^0(y)\}\approx-\left[
  \gamma^{ij}(x) \tilde{R}_j(x) \frac{\partial}{\partial x^i}
  \delta^3(x-y) - \gamma^{ij}(y) \tilde{R}_j(y)
  \frac{\partial}{\partial y^i} \delta^3(x-y) \right] \text{,} 
\end{equation}
which yields $\gamma_{ij}$ as the spatial HKT metric and $g_{\mu \nu}$
as the spacetime metric. Hence, both $g_{\mu \nu}$ and $f_{\mu \nu}$
can arise as HKT metrics, consistent with the general structure of the
theory.\footnote{What determines the metric is not so much
  the field redefinition \eqref{ndef}, but rather what shifts are
  retained as Lagrange multipliers. If $N^i$ are retained the
  HKT-metric is $g_{\mu \nu}$, whereas $L^i$ lead to $f_{\mu \nu}$.}
This is also consistent with the nature of ghost free matter
couplings, whereby, a certain type of matter can only couple minimally
to one of the metrics.

The nonuniqueness of this metric is easy to understand. The
observation made by HKT was that whenever the Hamiltonian can be
expressed as $\mathcal{H} = - M \tilde{R}^0 - M^i \tilde{R}_i$, with
first class constraints $\tilde{R}^0$ and $\tilde{R}_i$ satisfying the
covariance algebra, then a metric can be read off from the
$\{\tilde{R}^0(x),\tilde{R}^0(y) \}$ bracket as described
above. However, in a theory where diffeomorphisms act on different
sectors, there may exist multiple ways of selecting the first class
constraints (and the Lagrange multipliers) each leading to a different
choice of the HKT metric. In bimetric theory there is a further
restriction. If, for some choice of variables, the HKT metric involves
a combination of $g$ and $f$ metrics, then that combination must
transform as a rank-2 tensor whenever $g$ and $f$ are transformed
under coordinate transformations.  For example, if we follow the steps
leading to the Hamiltonian \eqref{fHamiltonian}, but then use $N=WL$
to eliminate $L$ in \eqref{bi-Hamiltonian}, we obtain (with $\hat{\hat
  R}^0=\hat{R}^0/W$),
\begin{equation} \label{f'Hamiltonian}
\mathcal{H} = - N \hat{\hat R}^0 - L^i \tilde{R}_i \text{.}
\end{equation}
However, a covariant metric with lapse $N$ and shift $L^i$ does not
exist. Indeed, the algebra of $\hat{\hat R}^0$ and $\tilde{R}_i$ is
not of the canonical form \eqref{GRalgebra}. Two different
redefinitions of the $\hat{\hat R}^0$ that restores the canonical form
of the algebra will lead to the HKT metric being $f_{\mu\nu}$ or
$g_{\mu\nu}$.

It may be possible to choose other variables than the ones described
above. For example one may use any linear combination of the shifts,
$M^i = aN^i + bL^i$ (with $a \neq -b$) as Lagrange multipliers. Using
stability conditions of the constraints one would then put the
Hamiltonian in the form
\begin{equation}
\mathcal{H} = - M^\mu R_\mu \text{,}
\end{equation}
If the $R_\mu$ algebra has the canonical form, it will be possible to
identify a new HKT metric with $M^0$ and $M^i$ as the lapse and
shift. This metric may be a complicated function of the phase space
variables that depends on $a$ and $b$. The metric identified in
\cite{krasnov} could be an example of this.

Note that, in any case, the physical gravitational metric is the one
that minimally couples to matter, say $g_{\mu\nu}$, as dictated by
absence of the ghost \cite{bimetric}.

\section*{Acknowledgments}

It is a pleasure to thank Ingemar Bengtsson and Mikica Kocic for fruitful discussions. This work was supported by a grant from the Swedish Research Council.

\appendix



\section{Determining the Lagrange multipliers} \label{det_Lagrange}

In massive gravity, the requirement that $\mathcal{C}_{(2)}\approx 0 $
is preserved under time evolution, gives an equation that determines
the Lagrange multiplier $N$ \cite{conf}. In bimetric gravity, the
corresponding equation depends on all five Lagrange multipliers $N,
L,$ and $L_i$,
\begin{align} \label{C2cond}
&\frac{\mathrm{d}}{\mathrm{d}t} \mathcal{C}_{(2)}(x) =
  \{\mathcal{C}_{(2)}(x) , H \} \notag \\  
= & -\int \mathrm{d}y \left[ L^i(y)
  \{\mathcal{C}_{(2)}(x),\tilde{R}_i(y) \} + L(y)\{\mathcal{C}_{(2)}(x),
  \tilde{R}^0(y) \} + N(y)\{\mathcal{C}_{(2)}(x),\mathcal{C}(y) \}\right] \approx 0 \text{,} 
\end{align}
However, now four of the multipliers are determined by gauge fixing
general coordinate transformations, analogous to the procedure 
in \cite{ADM}. Then \eqref{C2cond} becomes an equation for the fifth
multiplier.  

Here, we demonstrate this procedure explicitly in the linearized
$\beta_1$-model: using general covariance to determine $L$ and $L^i$,
\eqref{C2cond} becomes an equation for $N$. We start with the
equations of motion for $\gamma_{ij}$, $\f_{ij}$, $\pi^{ij}$, and
$p^{ij}$.  Since the interaction terms in the action \eqref{bi-action}
do not depend on $\pi^{ij}$, varying with respect to it gives the
same equation for $\gamma_{ij}$ as in general relativity,
\begin{equation}
\partial_t \gamma_{ij} = \frac{2N}{M_g^2 \sqrt{\det \gamma}} \left( \pi_{ij} - \frac{1}{2} \gamma_{ij} \pi \right) +  \nabla_i N_j + \nabla_j N_i \text{,}
\end{equation}
where, $N_i$ are given by equation \eqref{ndef} . Similarly, varying
with respect to $p^{ij}$ leads to equation \eqref{3feom} for
$\f_{ij}$.  However, since the interaction terms of \eqref{bi-action}
depend on $\gamma_{ij}$, the equation for $\pi^{ij}$ will have extra
terms compared to the corresponding equation in general relativity.
The extra terms can be computed using \eqref{Sdef}, \eqref{SV} and 
\eqref{Umn-def}, and the equation becomes,
\begin{align}
\partial_t \pi^{ij} =& -M_g^2 N \sqrt{\det \gamma} \left( \R^{ij (g)} - \frac{1}{2} \gamma^{ij} \R^{(g)} \right) + \frac{N}{2 M_g^2 \sqrt{\det \gamma}} \gamma^{ij} \left( \pi^{mn} \pi_{mn} - \frac{1}{2} \pi^2 \right) \notag \\
& -\frac{2N}{M_g^2 \sqrt{\det \gamma}} \left( \pi^{ik} {\pi_k}^j - \frac{1}{2} \pi \pi^{ij} \right) + M_g^2 \sqrt{\det \gamma} \left( \nabla^i \nabla^j N - \gamma^{ij} \nabla^k \nabla_k N \right) \notag \\
&+ \sqrt{\det \gamma} \left[ \nabla_k \left( \frac{N^k
   \pi^{ij}}{\sqrt{\det \gamma}} \right) -\nabla_k \left( \frac{N^i
    \pi^{jk}}{\sqrt{\det \gamma}}  \right) - \nabla_k \left( \frac{N^j \pi^{ik}}{\sqrt{\det \gamma}} \right) \right] \notag \\
&+ m^4 \sqrt{\det \gamma} \left[ N \left( V \gamma^{ij} - \bar{V}^{ij} \right) + L U^{ij} \right] \text{.}
\end{align}
In the above, we express the $N^i$  in terms of $n^i$ given by
equation \eqref{ndef}. Analogously, using equations \eqref{Zdef} and
\eqref{ZW}, the equation for $p^{ij}$ is (for general $\beta_n$), 
\begin{align}
\partial_t p^{ij} =& -M_f^2 L \sqrt{\det \f} \left( \R^{ij (f)} - \frac{1}{2} \f^{ij} \R^{(f)} \right) + \frac{L}{2 M_f^2 \sqrt{\det \f}} \f^{ij} \left( p^{mn} p_{mn} - \frac{1}{2} p^2 \right) \notag \\
& -\frac{2L}{M_f^2 \sqrt{\det \f}} \left( p^{ik} {p_k}^j - \frac{1}{2} p p^{ij} \right) + M_f^2 \sqrt{\det \f} \left( \bar{\nabla}^i \bar{\nabla}^j L - \f^{ij} \bar{\nabla}^k \bar{\nabla}_k L \right) \notag \\
&+ \sqrt{\det \f} \left[ \bar{\nabla}_k \left( \frac{L^k p^{ij}}{\sqrt{\det \f}} \right) -  \bar{\nabla}_k \left( \frac{L^i p^{jk}}{\sqrt{\det \f}} \right) - \bar{\nabla}_k \left( \frac{L^j p^{ik}}{\sqrt{\det \f}} \right) \right] \notag \\
&+ m^4 \sqrt{\det \gamma} \left( N \bar{W}^{ij} + L \tilde{U}^{ij} \right) \text{,}
\end{align}
Here, in analogy with \eqref{Umn-def}, $\tilde{U}^{mn}$ is defined as 
\begin{equation} \label{Utildedef}
\tilde{U}^{mn} \equiv 2 \frac{\partial U'}{\partial \f_{mn}} \text{,}
\end{equation}

We now focus on the $\beta_1$-model by setting
$\beta_2=\beta_3=0$. The equations will admit flat background
solutions $g_{\mu \nu}=f_{\mu \nu}=\eta_{\mu \nu}$ if $\beta_0=-3
\beta_1$, and $\beta_4=-\beta_1$. which we also assume \cite{HSS}.  To
linearize the theory around the flat background, we consider the small
perturbations,
\begin{equation}
N = 1 + \delta N \text{,} \qquad  \gamma_{ij} = \delta_{ij} +
\hat{\gamma}_{ij} \text{,} \qquad L = 1 + \delta L \text{,} \qquad
L_i = \delta L_i \text{,} \qquad \f_{ij} = \delta_{ij} + \fhat_{ij}
\text{,} 
\end{equation}
with the boundary condition that all the perturbations vanish at infinity.
A symmetric tensor perturbation admits the following linear orthogonal
decomposition \cite{ADM}, 
\begin{equation} \label{ortdec}
a_{ij} = a_{ij}^{TT} + a_{ij}^T + \partial_i a_j + \partial_j a_i \text{.}
\end{equation}
The components are defined in terms of $a^T$, 
\begin{equation} \label{aTdef}
a^T = a - \frac{1}{\nabla^2} \partial^i \partial^j a_{ij} \text{,}
\end{equation}
and are given by,
\begin{align} \label{ortdeccomp}
&a_i = \frac{1}{\nabla^2} \left( \partial^j a_{ij} - \frac{1}{2} \frac{1}{\nabla^2} \partial_i \partial^j \partial^k a_{jk} \right) \text{,} \notag \\
&a_{ij}^{T} = \frac{1}{2} \left( \delta_{ij} a^T - \frac{1}{\nabla^2} \partial_i \partial_j a^T \right) \text{,} \\
&a_{ij}^{TT}  = a_{ij} - a_{ij}^T - \partial_i a_j - \partial_j a_i \text{.} \notag
\end{align}
Here, $a$ is the trace of $a_{ij}$, defined as $a = a_{ij}
\delta^{ij}$, and $1/\nabla^2$ is the inverse of the flat space
Laplacian. 

In the $\beta_1$-model, the linearized equations of motion become
\begin{equation} \label{lineargeom}
\partial_t \hat{\gamma}_{ij} = \frac{2}{M_g^2} \left( \pi_{ij} - \frac{1}{2} \delta_{ij} \pi \right) + \partial_i L_j + \partial_j L_i + \frac{2}{m^4 \beta_1} \left( \partial_i \partial^k \pi_{jk} + \partial_j \partial^k \pi_{ik} \right) \text{,}
\end{equation}
\begin{equation} \label{linear3feom}
\partial_t \fhat_{ij} = \frac{2}{M_f^2} \left( p_{ij} - \frac{1}{2} \delta_{ij} p \right) + \partial_i L_j + \partial_j L_i \text{,}
\end{equation}
\begin{align}
\partial_t \pi^{ij} = &- \frac{M_g^2}{2} \left( \partial_k \partial^i \hat{\gamma}^{jk} + \partial _k \partial^j \hat{\gamma}^{ik} - \partial^i \partial^j \hat{\gamma} - \nabla^2 \hat{\gamma}^{ij} \right) \notag \\
& + \frac{M_g^2}{2} \delta^{ij} \left( \partial^k \partial^l \hat{\gamma}_{kl} - \nabla^2 \hat{\gamma} \right) + M_g^2 \left( \partial^i \partial^j \delta N - \delta^{ij} \nabla^2 \delta N \right) \notag \\
& - m^4 \beta_1 \left( \frac{1}{2} \fhat^{ij} - \frac{1}{2} \hat{\gamma}^{ij} -\delta^{ij} \frac{\fhat}{2} + \delta^{ij} \frac{\hat{\gamma}}{2} - \delta^{ij} \delta L + \delta^{ij} \delta N    \right) \text{,}
\end{align}
\begin{align} \label{linearpeom}
\partial_t p^{ij} = &- \frac{M_f^2}{2} \left( \partial_k \partial^i \fhat^{jk} + \partial _k \partial^j \fhat^{ik} - \partial^i \partial^j \fhat - \nabla^2 \fhat^{ij} \right) \notag \\
& + \frac{M_f^2}{2} \delta^{ij} \left( \partial^k \partial^l \fhat_{kl} - \nabla^2 \fhat \right) + M_f^2 \left( \partial^i \partial^j \delta L - \delta^{ij} \nabla^2 \delta L \right) \notag \\
& + m^4 \beta_1 \left( \frac{1}{2} \fhat^{ij} - \frac{1}{2} \hat{\gamma}^{ij} -\delta^{ij} \frac{\fhat}{2} + \delta^{ij} \frac{\hat{\gamma}}{2} - \delta^{ij} \delta L + \delta^{ij} \delta N    \right) \text{,}
\end{align}
where, in the derivation of \ref{lineargeom}, we have used
\begin{equation}
n^i = \frac{\f^{ij} R^{(g)}_j}{\sqrt{4m^8 \beta_1^2 \det \gamma + R^{(g)}_k \f^{kl} R^{(g)}_l}} \text{,}
\end{equation}
which is the solution to \ref{neqom} in the $\beta_1$-model
\cite{fra}. In addition, the constraints $\tilde{R}^0 = 0$ and
$\mathcal{C}=0$, with $\tilde{R}^0$ and $\mathcal{C}$ given by
\eqref{R0tilde_def} and \eqref{bi-C}, are linearized to  
\begin{equation} \label{linearR0}
M_f^2 \left( \partial^i \partial^j \fhat_{ij} - \nabla^2 \fhat \right) + m^4 \beta_1 \left(\hat{\gamma} - \fhat \right) = 0 \text{,}
\end{equation}
\begin{equation} \label{linearC}
M_g^2 \left( \partial^i \partial^j \hat{\gamma}_{ij} - \nabla^2 \hat{\gamma} \right) + m^4 \beta_1 \left(\fhat - \hat{\gamma} \right) = 0 \text{.}
\end{equation}
In order to determine $L$ and $L^i$, coordinate conditions must be
imposed. Here we follow a prescription outlined in \cite{HK} and 
impose coordinate conditions with respect to the composite metric,
\begin{equation}
h_{\mu \nu} = g_{\mu \lambda} {\left( \sqrt{g^{-1} f} \right)^\lambda}_\nu \text{,}
\end{equation}
which results in simpler equations. The null cone of $h_{\mu \nu}$
always encloses the intersection of the null cones of $g_{\mu\nu}$ and
$f_{\mu \nu}$ \cite{HK}. Carrying out a $3+1$ decomposition of
$h_{\mu\nu}$ around a flat background,
\begin{equation}
  H \equiv \left(-h^{00} \right)^{-1/2}=1 + \delta H   \text{,}
  \qquad
  H_i \equiv h_{0 i}=\delta H_i \text{,}
  \qquad
  \h_{ij} \equiv h_{ij}=\delta_{ij}+\hat{h}_{ij} \text{,} 
\end{equation}
we obtain, at the linear level,
\begin{equation}
\hat{h}_{ij} = \frac{1}{2} \left( \hat{\gamma}_{ij} + \fhat_{ij} \right) \text{.}
\end{equation}
The canonical momentum of $\h_{ij}$ is given by
\begin{equation}
\pi_h^{ij} = 2 \left( \pi^{ij} + p^{ij} \right) \text{.}
\end{equation}

We can now use this to choose a coordinate condition to fix a time
direction, as well as a spacelike hypersurface. In analogy with
\cite{ADM}, we choose 
\begin{equation} \label{pihcond}
\nabla^2 \pi_h - \partial_i \partial_j \pi_h^{ij} = 0 \text{,}
\end{equation}
which ensures that the time direction lies inside the null cones of
both $g_{\mu \nu}$ and $f_{\mu \nu}$. We must also choose coordinate
conditions on the hypersurface. In principle the condition
$\partial^j \hat{h}_{ij} = 0$ could be used, to extend the analogy to
\cite{ADM}. However, the equations turns out to be simpler if the
condition   
\begin{equation} \label{fcond}
\partial^j \fhat_{ij} = 0 \text{.}
\end{equation}
is used instead. The coordinate conditions \eqref{pihcond} and
\eqref{fcond} can be rewritten as 
\begin{equation} \label{pihTcond}
\pi_h^T = 0 \Rightarrow \pi^T + p^T = 0 \text{,}
\end{equation}
\begin{equation} \label{ficond}
\fhat_i = 0 \text{,}
\end{equation}
using the orthogonal decomposition \eqref{ortdec} \cite{ADMIVC}. When
rewriting them this way, one makes use of 
\begin{equation}
\frac{1}{\nabla^2} 0 = 0 \text{,}
\end{equation}
which holds for the vanishing boundary conditions at infinity
\cite{Regge}. Let us now define,
\begin{equation}
P_{ij} \equiv p_{ij} - \frac{1}{2} \delta_{ij} p \text{,}
\end{equation}
and use $P_{ij}$ to express \eqref{linear3feom} as
\begin{equation}
\partial_t \fhat_{ij} = \frac{2}{M_f^2} P_{ij} + \partial_i L_j + \partial_j L_i \text{.}
\end{equation}
From the definition of the vector component, $a_i$, in \eqref{ortdeccomp}, it follows that if $a_{ij} = \partial_i L_j + \partial_j L_i$ then $a_i = L_i$. This leads us to the equation
\begin{equation} \label{fieom}
\partial_t \fhat_i = \frac{2}{M_f^2} P_i + L_i \text{,}
\end{equation}
where
\begin{align}
P_i = \frac{1}{\nabla^2} \left[ \partial^j p_{ij} - \frac{1}{4} \partial_i p - \frac{1}{2} \partial_i \frac{1}{\nabla^2} \partial^j \partial^k p_{jk} \right] \text{.}
\end{align}
This, together with \eqref{ficond}, means \eqref{fieom} can be rewritten as
\begin{equation}
L_i = \frac{2}{M_f^2} \frac{1}{\nabla^2} \left( \partial^j p_{ij} - \frac{1}{4} \partial_i p - \frac{1}{2} \partial_i \frac{1}{\nabla^2} \partial^j \partial^k p_{jk} \right) \text{.}
\end{equation}
The Lagrange multipliers $L^i$ are therefore determined in terms of
the dynamical variables. Note that this is consistent with the
boundary conditions, since both $L_i$ and $p^{ij}$ vanish at
infinity. 

We now turn our attention to $L$. By imposing the linearized
constraint \eqref{linearR0}, the linearized $p^{ij}$ equation of
motion, \eqref{linearpeom}, can be simplified to 
\begin{align}
\partial_t p^{ij} = &- \frac{M_f^2}{2} \left( \partial_k \partial^i \fhat^{jk} + \partial _k \partial^j \fhat^{ik} - \partial^i \partial^j \fhat - \nabla^2 \fhat^{ij} \right) \notag \\
& +  M_f^2 \left( \partial^i \partial^j \delta L - \delta^{ij} \nabla^2 \delta L \right)  + m^4 \beta_1 \left( \delta^{ij} \delta N - \delta^{ij} \delta L - \frac{1}{2} \hat{\gamma}^{ij} + \frac{1}{2} \fhat^{ij}   \right) \text{.}
\end{align}
From this, we derive the equation of motion for $p^T$ by computing the
trace of the transverse part of each term on the right hand side,
using \eqref{aTdef}. This gives the result 
\begin{align}
\partial_t p^T =& - \frac{M_f^2}{2} \left( \partial^i \partial^j \fhat_{ij} - \nabla^2 \fhat \right)  - 2 M_f^2 \nabla^2 \delta L \notag \\
& + 2m^4 \beta_1 \left( \delta N - \delta L \right)  - \frac{m^4 }{2} \beta_1 \left( \hat{\gamma}^T - \fhat^T \right) \text{,}
\end{align}
Similarly, using \eqref{aTdef} and \eqref{linearC}, the equation of motion for $\pi^T$ becomes
\begin{align}
\partial_t \pi^T =& - \frac{M_g^2}{2} \left( \partial^i \partial^j \hat{\gamma}_{ij} - \nabla^2 \hat{\gamma} \right)  - 2 M_g^2 \nabla^2 \delta N \notag \\
& - 2m^4 \beta_1 \left( \delta N - \delta L \right)  + \frac{m^4 }{2} \beta_1 \left( \hat{\gamma}^T - \fhat^T \right) \text{.}
\end{align}
Adding these equations yields
\begin{align}
\partial_t \left(\pi^T + p^T \right) =& - \frac{M_g^2}{2} \left( \partial^i \partial^j \hat{\gamma}_{ij} - \nabla^2 \hat{\gamma} \right) - \frac{M_f^2}{2} \left( \partial^i \partial^j \fhat_{ij} - \nabla^2 \fhat \right) \notag \\
& - 2 M_g^2 \nabla^2 \delta N - 2 M_f^2 \nabla^2 \delta L \text{,}
\end{align}
which, using the coordinate condition \eqref{pihTcond} as well as the constraints \eqref{linearR0} and \eqref{linearC}, reduces to
\begin{equation}
2 M_g^2 \nabla^2 \delta N + 2 M_f^2 \nabla^2 \delta L = 0 \text{,}
\end{equation}
which in turn simplifies to
\begin{equation} \label{deltaLeq}
M_g^2 \delta N +  M_f^2 \delta L = 0 \text{.}
\end{equation}

In order to solve for $\delta L$ and $\delta N$ in terms of the
dynamical variables, the linearized version of equation \eqref{C2cond}
is needed. This equation will now be derived. The linearized version
of the secondary constraint, equation \eqref{bimetricC2_final_form},
is 
\begin{equation} \label{linearC2}
\mathcal{C}_{(2)} = -2 \partial_i \partial_j \pi^{ij} + \beta_1 \frac{m^4}{M_g^2} \pi - \beta_1 \frac{m^4}{M_f^2} p \text{,}
\end{equation}
and the quadratic Hamiltonian is
\begin{align} \label{quadH}
H = - \int \diff^3 y & \Bigg[ \frac{M_g^2}{2} \left( - \frac{1}{2} \partial_k \ghat_{ij} \partial^k \ghat^{ij} + \partial_i \ghat_{jk} \partial^j \ghat^{ik} - \partial_i \ghat^{ij} \partial_j \ghat + \frac{1}{2} \partial_k \ghat \partial^k \ghat \right) \notag \\
& + \frac{M_f^2}{2} \left( - \frac{1}{2} \partial_k \fhat_{ij} \partial^k \fhat^{ij} + \partial_i \fhat_{jk} \partial^j \fhat^{ik} - \partial_i \fhat^{ij} \partial_j \fhat + \frac{1}{2} \partial_k \fhat \partial^k \fhat \right) \notag \\
&+ M_g^2 \delta N \left( \partial_i \partial_j \ghat^{ij} - \nabla^2 \ghat \right) + M_f^2 \delta L \left( \partial_i \partial_j \fhat^{ij} - \nabla^2 \fhat \right)  \notag \\
&+  \frac{m^4 \beta_1}{4} \left( \ghat^{ij} \ghat_{ij} - \ghat^2 + \fhat^{ij} \fhat_{ij} - \fhat^2 - 2 \ghat^{ij} \fhat_{ij} + 2 \ghat \fhat \right) \notag \\
& + m^4 \beta_1 \left( \fhat \left( \delta N - \delta L \right) - \ghat \left( \delta N - \delta L \right) \right) + \hat{\mathcal{H}}( \pi , p ) \Bigg] \text{,}
\end{align}
where $\hat{\mathcal{H}}( \pi , p )$ is a function of $\pi^{ij}$ and
$p^{ij}$, the exact form of which will not matter in the calculations
to come. Since the linearized $\mathcal{C}_{(2)}$ only depends on the
momenta, the the linearized version of equation \eqref{C2cond} can be
written 
\begin{equation} \label{C2dotbracket}
\dot{\mathcal{C}}_{(2)}(x) = \{ \C(x) ,H \} = - \int \diff^3 z \left( \frac{\delta \C(x)}{\delta \pi^{mn}(z)} \frac{\delta H}{\delta \ghat_{mn}(z)} + \frac{\delta \C(x)}{\delta p^{mn}(z)} \frac{\delta H}{\delta \fhat_{mn}(z)} \right) \text{,}
\end{equation}
where $\C$ and $H$ are given by equation \eqref{linearC2} and
\eqref{quadH}. From these equations it follows that 
\begin{equation}
\frac{\delta \C(x)}{\delta \pi^{mn}(z)} = - 2 \frac{\partial}{\partial x^m} \frac{\partial}{\partial x^n} \delta^3(x-z) + \beta_1 \frac{m^4}{M_g^2} \delta_{mn} \delta^3(x-z) \text{,}
\end{equation}
\begin{equation}
\frac{\delta \C(x)}{\delta p^{mn}(z)} =  - \beta_1 \frac{m^4}{M_f^2} \delta_{mn} \delta^3(x-z) \text{,}
\end{equation}
\begin{align}
\frac{\delta H}{\delta \ghat(z)} = - & \bigg[ \frac{M_g^2}{2} \left( \nabla^2 \ghat^{mn} - \partial_i \partial^m \ghat^{in} - \partial_i \partial^n \ghat^{im} + \partial^m \partial^n \ghat + \delta^{mn} \partial_i \partial_j \ghat^{ij} - \delta^{mn} \nabla^2 \ghat \right) \notag \\
& + M_g^2 \left( \partial^m \partial^n \delta N - \delta^{mn} \nabla^2 \delta N \right) \notag \\
& + m^4 \beta_1 \left( \frac{1}{2} \left( \ghat^{mn} - \delta^{mn} \ghat - \fhat^{mn} + \delta^{mn} \fhat \right) - \delta^{mn} \left( \delta N - \delta L \right) \right) \bigg]
\end{align}
\begin{align}
\frac{\delta H}{\delta \fhat(z)} = - & \bigg[ \frac{M_f^2}{2} \left( \nabla^2 \fhat^{mn} - \partial_i \partial^m \fhat^{in} - \partial_i \partial^n \fhat^{im} + \partial^m \partial^n \fhat + \delta^{mn} \partial_i \partial_j \fhat^{ij} - \delta^{mn} \nabla^2 \fhat \right) \notag \\
& + M_f^2 \left( \partial^m \partial^n \delta L - \delta^{mn} \nabla^2 \delta L \right) \notag \\
& + m^4 \beta_1 \left( \frac{1}{2} \left( \fhat^{mn} - \delta^{mn} \fhat - \ghat^{mn} + \delta^{mn} \ghat \right) + \delta^{mn} \left( \delta N - \delta L \right) \right) \bigg] \text{.}
\end{align}
Inserting these into equation \eqref{C2dotbracket} yields the result
\begin{align}
\dot{\mathcal{C}}_{(2)} = m^4 \beta_1 & \Bigg[ \frac{1}{2} \left( \partial_i \partial_j \fhat^{ij} - \nabla^2 \fhat \right) - \frac{1}{2} \left( \partial_i \partial_j \ghat^{ij} - \nabla^2 \ghat \right) \notag \\
& + m^4 \beta_1 \left( \fhat - \ghat - 3 \left( \delta N - \delta L \right) \right) \left( \frac{1}{M_g^2} + \frac{1}{M_f^2} \right) \Bigg] \text{,}
\end{align}
which, after imposing the constraints \eqref{linearR0} and \eqref{linearC}, simplifies to
\begin{equation} \label{linearC2dot}
\dot{\mathcal{C}}_{(2)} \approx 3 m^8 \beta_1^2 \left( \frac{1}{M_g^2} + \frac{1}{M_f^2} \right) \left( \frac{1}{2} \left( \fhat - \ghat \right) + \delta L - \delta N \right) \text{.}
\end{equation}
$\dot{\mathcal{C}}_{(2)} \approx 0$ then implies
\begin{equation}
\delta N - \delta L = \frac{1}{2} \left( \fhat - \ghat \right) \text{,}
\end{equation}
which, together with equation \eqref{deltaLeq}, can be solved for
$\delta L$ and $\delta N$. The result is 
\begin{equation}
\delta L = - \frac{M_g^2}{2 \left( M_g^2 + M_f^2 \right)} \left( \fhat - \ghat \right) \text{,}
\end{equation}
\begin{equation}
\delta N = \frac{M_f^2}{2 \left( M_g^2 + M_f^2 \right)} \left( \fhat - \ghat \right) \text{.}
\end{equation}
This shows that all Lagrange multipliers in bimetric gravity can be
determined, using the consistency condition that $\mathcal{C}_{(2)}$
is preserved in time, as well as suitable coordinate
conditions. This procedure can be generalized to the non-linear case,
in a manner similar to that described in \cite{ADM}. 


\section{Computation of the Poisson brackets} \label{appendix_brackets}

Here we compute the Poisson brackets of the constraints
$\tilde{R}^0$, $\tilde{R}_i$, $\cal C$, and $\C$. A computation
of the algebra of constraints of general relativity \eqref{GRalgebra}
can be found in the appendix of \cite{spatcov} to which we refer for
the relevant details. In particular, we reproduce the following
results that will be useful in the computation of brackets in bimetric
theory.  To systematically deal with the derivatives of the delta
function that arise in the calculations one can introduce smoothing
functions and define,
\begin{equation} \label{Agdef}
A_H^{(g)} \equiv \int \mathrm{d}^3x a^0(x) R^{0 (g)}(x) \text{,}
\qquad A^{(g)} \equiv \int \mathrm{d}^3x a^i(x) R_i^{(g)}(x) \text{,} 
\end{equation}
\begin{equation} \label{Bgdef}
B_H^{(g)} \equiv \int \mathrm{d}^3y b^0(y) R^{0 (g)}(y) \text{,}
\qquad B^{(g)} \equiv \int \mathrm{d}^3y b^i(y) R_i^{(g)}(y) \text{,} 
\end{equation}
where $a^0$, $a^i$, $b^0$ and $b^i$ are time independent smoothing
functions. These can be used to compute the brackets of the constraints
involving the metric $g_{\mu\nu}$, since, for example, 
\begin{equation}
\{A^{(g)},B^{(g)} \} = \int \mathrm{d}^3x \int \mathrm{d}^3y a^i(x)
b^i(y) \{ R_i^{(g)}(x),R_i^{(g)}(y) \} \text{.} 
\end{equation}
Then, one can obtain the following intermediate relations,  
\begin{equation} \label{dA/dgamma}
\frac{\delta A^{(g)}}{\delta \gamma_{mn}} = \sqrt{\det \gamma}
\nabla_i \left( a^i \frac{\pi^{mn}}{\sqrt{\det \gamma}} \right) -
\pi^{mi} \nabla_i a^n - \pi^{ni} \nabla_i a^m \text{,} 
\end{equation}
\begin{equation} \label{dA/dpi}
\frac{\delta A^{(g)}}{\delta \pi^{mn}} = - \left( \gamma_{nj} \nabla_m
a^j + \gamma_{mj} \nabla_n a^j \right) \text{,} 
\end{equation}
\begin{equation} \label{dAH/dpi}
\frac{\delta A_H^{(g)}}{\delta \pi^{mn}} = - \frac{a^0}{M_g^2
  \sqrt{\det \gamma}} \left( 2 \pi_{mn} - \gamma_{mn} {\pi^k}_k
\right) \text{,} 
\end{equation}
\begin{equation} \label{ABbracket_g}
\{A^{(g)},B^{(g)} \} =  - \int \mathrm{d}^3z \left( a^i R_j^{(g)}
\nabla_i b^j - b^j R_i^{(g)} \nabla_j a^i \right) \text{,} 
\end{equation}
\begin{equation} \label{AHBbracket_g}
\{A_H^{(g)},B^{(g)} \} = - \int \mathrm{d}^3z a^0 \nabla_i \left(b^i
R^{0 (g)} \right) \text{,} 
\end{equation}
\begin{equation}\label{AHBHbracket_g} 
\{A_H^{(g)},B_H^{(g)}\} = - \int \mathrm{d}^3z \left( a^0 R^{i (g)}
\nabla_i b^0 - b^0 R^{i (g)} \nabla_i a^0 \right) \text{.} 
\end{equation}
Similarly, in analogy with \eqref{Agdef} and \eqref{Bgdef}, for the
$f_{\mu\nu}$ metric one can define, 
\begin{equation} \label{Afdef}
A_H^{(f)} \equiv \int \mathrm{d}^3x a^0(x) R^{0 (f)}(x) \text{,} \qquad A^{(f)} \equiv \int \mathrm{d}^3x a^i(x) R_i^{(f)}(x) \text{,}
\end{equation}
\begin{equation} \label{Bfdef}
B_H^{(f)} \equiv \int \mathrm{d}^3y b^0(y) R^{0 (f)}(y) \text{,} \qquad B^{(f)} \equiv \int \mathrm{d}^3y b^i(y) R_i^{(f)}(y) \text{,}
\end{equation}
which satisfy the same relations as those for the $g_{\mu\nu}$
metric above.

\subsection{Evaluation of $\{ \tilde{R}^0(x),\tilde{R}_i(y) \}$}

Corresponding to the bimetric constraints $\tilde{R}^0$
\eqref{R0tilde_def} and $\tilde{R}_i$ \eqref{Ritilde_def}, let us
introduce the smeared functions,  
\begin{equation} \label{FGdef}
F \equiv \int \mathrm{d}^3x f(x) \tilde{R}^0(x) \text{,} \qquad G
\equiv \int \mathrm{d}^3y g^i(y) \tilde{R}_i(y) \text{,} 
\end{equation}
where, $f(x)$ and $g^i(y)$ are time independent smoothing
functions. It follows that 
\begin{equation} \label{FGbracket}
\{F,G\} = \int \mathrm{d}^3x \int \mathrm{d}^3y f(x) g^i(y) \{
\tilde{R}^0(x),\tilde{R}_i(y) \} \text{.} 
\end{equation}
From the definition of the Poisson bracket, this can also be written as
\begin{align}
\{F,G \} =  \{F,G\}_g + \{F,G\}_f  &= \int \mathrm{d}^3 z \left(
\frac{\delta F}{\delta \gamma_{mn}(z)} \frac{\delta G}{\delta
  \pi^{mn}(z)} - \frac{\delta F}{\delta \pi^{mn}(z)} \frac{\delta
  G}{\delta \gamma_{mn}(z)} \right) \notag \\  
&+ \int \mathrm{d}^3 z \left( \frac{\delta F}{\delta \f_{mn}(z)}
\frac{\delta G}{\delta p^{mn}(z)} - \frac{\delta F}{\delta p^{mn}(z)}
\frac{\delta G}{\delta \f_{mn}(z)} \right) \text{.} 
\end{align}
The variations of $G$ with respect to $\gamma_{ij}$ and $\pi^{ij}$ are
given by \eqref{dA/dgamma} and \eqref{dA/dpi}, 
\begin{equation}
\frac{\delta G}{\delta \gamma_{mn}} = \sqrt{\det \gamma} \nabla_i
\left( g^i \frac{\pi^{mn}}{\sqrt{\det \gamma}} \right) - \pi^{mi}
\nabla_i g^n - \pi^{ni} \nabla_i g^m \text{,} 
\end{equation}
\begin{equation}  \label{dG/dpi}
\frac{\delta G}{\delta \pi^{mn}} = - \left( \gamma_{nj} \nabla_m g^j +
\gamma_{mj} \nabla_n g^j \right) \text{.} 
\end{equation}
The variations with respect to $\f_{ij}$ and $p^{ij}$ are given by 
the analogue of \eqref{dA/dgamma} and \eqref{dA/dpi} for $f_{\mu\nu}$, 
\begin{equation}
\frac{\delta G}{\delta \f_{mn}} = \sqrt{\det \f}  \bar{\nabla}_i
\left( g^i \frac{p^{mn}}{\sqrt{\det \f}} \right) - p^{mi}
\bar{\nabla}_i g^n - p^{ni} \bar{\nabla}_i g^m \text{,} 
\end{equation}
\begin{equation} \label{dG/dp}
\frac{\delta G}{\delta p^{mn}} = - \left( \f_{nj} \bar{\nabla}_m g^j +
\f_{mj} \bar{\nabla}_n g^j \right) \text{.} 
\end{equation}
From equation \eqref{dR0/dn}, it follows that when computing the
variations of $F$, we need not consider the implicit dependence on the
dynamical variables through $n^i$, but can in fact keep $n^i$
fixed. It is therefore useful to introduce the notation, 
\begin{equation}
f^i = f n^i \text{,}
\end{equation}
and define the quantities,
\begin{equation}\label{BFDF}
B_F \equiv \int \mathrm{d}^3x f(x) R^{0 (f)}(x) \text{,} \qquad D_F
\equiv \int \mathrm{d}^3x f^i(x) R_i^{(g)}(x) \text{.} 
\end{equation}
From these definitions, and \eqref{R0tilde_def}, it follows that, 
\begin{align} \label{dF/dgamma}
&\frac{\delta F}{\delta \gamma_{mn}} = \frac{\delta D_F}{\delta
  \gamma_{mn}} + 2m^4 f \frac{\partial \left( \sqrt{ \det \gamma} U'
  \right) }{\partial \gamma_{mn}} \text{,} 
\qquad\qquad 
\frac{\delta F}{\delta \pi^{mn}} = \frac{\delta D_F}{\delta \pi^{mn}} \text{,}
\\
\label{dF/df}
&\frac{\delta F}{\delta \f_{mn}} = \frac{\delta B_F}{\delta \f_{mn}} +
2m^4 f \sqrt{ \det \gamma} \frac{\partial  U'  }{\partial \f_{mn}} \text{,} 
\qquad\qquad \,
\frac{\delta F}{\delta p^{mn}} = \frac{\delta B_F}{\delta p^{mn}} \text{.}
\end{align}
Then the bracket $\{F,G \}_g$ can then be written as,
\begin{align}
\{F,G \}_g = \{D_F,G \}_g + 2m^4 \int \mathrm{d}^3z f \frac{\partial
  \left( \sqrt{ \det \gamma} U' \right) }{\partial \gamma_{mn}}
\frac{\delta G}{\delta \pi^{mn}} \text{.} 
\end{align}
Since $n^i$ is kept fixed, the first term is similar in form to
\eqref{ABbracket_g}, and is therefore given by 
\begin{align}
\{D_F,G \}_g = - \int \mathrm{d}^3z \left( f^i R_j^{(g)} \nabla_i g^j
- g^j R_i^{(g)} \nabla_j f^i \right) \text{.} 
\end{align}
The second term is (since $\sqrt{\det \gamma} U'$ and $\sqrt{\det \gamma} U$ differ by a term independent of
$\gamma$),
\begin{align} \label{Umn-integral}
2m^4 \int \mathrm{d}^3z f \frac{\partial \left( \sqrt{ \det \gamma} U'
  \right) }{\partial \gamma_{mn}} \frac{\delta G}{\delta \pi^{mn}} 
= -2 m^4 \int \mathrm{d}^3z f \sqrt{\det \gamma} \left(U \nabla_j g^j  
+2 \frac{\partial U}{\partial \gamma_{mn}} \gamma_{nj} \nabla_m g^j 
\right)
\text{.} 
\end{align}
Now we turn our attention to $\{F,G \}_f$, which can be written as,
\begin{align}
\{F,G \}_f = \{B_F,G \}_f + 2m^4 \int \mathrm{d}^3z f \sqrt{ \det
  \gamma} \frac{\partial  U'  }{\partial \f_{mn}} \frac{\delta
  G}{\delta p^{mn}} \text{.} 
\end{align}
The first term is analogous to \eqref{AHBbracket_g}, and is therefore
given by 
\begin{equation}
\{B_F,G \}_f =-\int\mathrm{d}^3z f\partial_i\left(g^i R^{0 (f)} \right) \text{.}
\end{equation}
Here as have used
\begin{equation} \label{densityderivative}
\bar{\nabla}_i \left(g^i R^{0 (f)} \right) = \partial_i \left(g^i R^{0 (f)} \right) \text{,}
\end{equation}
which follows since $g^i$ is a vector and $R^{0 (f)}$ is a scalar
density\footnote{To derive the identity, one needs the formula for
  the covariant  derivative of a scalar density,   $\nabla_\alpha D =
  \partial_\alpha D - \tensor[]{\Gamma}{^\mu _\alpha _\mu} D$.}. The
second term becomes, 
\begin{equation}
2m^4 \int \mathrm{d}^3z f \sqrt{ \det \gamma} \frac{\partial  U'
}{\partial \f_{mn}} \frac{\delta G}{\delta p^{mn}} = -2m^4 \int
\mathrm{d}^3z f \sqrt{\det \gamma} \tilde{U}^{mn} \f_{nj}
\bar{\nabla}_m g^j \text{,} 
\end{equation}
where $\tilde{U}^{mn}$ is defined by \eqref{Utildedef}. Putting the
above results together, one obtains,
\begin{align} \label{FGbracket-cov}
\{F,G \} =& \int \mathrm{d}^3z \bigg[ - f \partial_i \left(g^i
  R^{0 (f)} \right) - f^i R_j^{(g)} \nabla_i g^j + g^j R_i^{(g)}
  \nabla_j f^i \notag \\ & -2 m^4 f \sqrt{\det \gamma} \left( U
  \nabla_i g^i + 2 \frac{\partial U}{\partial \gamma_{mn}} \gamma_{nj}
  \nabla_m g^j + \tilde{U}^{mn} \f_{nj} \bar{\nabla}_m g^j \right)
  \bigg] \text{.}
\end{align}

We now bring \eqref{FGbracket-cov} to the appropriate form by
rewriting the terms.  Using integration by parts, the third term can
be rewritten as, 
\begin{align} \label{fn_chainrule}
&\int \mathrm{d}^3z g^j R_i^{(g)} \nabla_j f^i = - \int \mathrm{d}^3z
  f \partial_i \left( g^i n^j R_j^{(g)} \right) + \int \mathrm{d}^3z f
  g^i R_j^{(g)} \nabla_i n^j \text{.} 
\end{align}
Using \eqref{U'-def} and \eqref{Utildedef}, it is possible to write,
\begin{equation}
f \sqrt{\det \gamma} \tilde{U}^{mn} \f_{nj} \bar{\nabla}_m g^j = 2 f
\sqrt{\det \gamma} \frac{\partial U}{\partial \f_{mn}} \f_{nj}
\bar{\nabla}_m g^j + \beta_4 f \sqrt{\det \f} \bar{\nabla}_i g^i
\text{.} 
\end{equation}
The last term in the above expression, together with the fourth term
in \eqref{FGbracket-cov}, can be rewritten in the following manner, 
\begin{align} \label{U_chainrule}
\int \mathrm{d}^3z f\left(\sqrt{\det \gamma} U \nabla_i g^i +
  \beta_4\sqrt{\det \f} \bar{\nabla}_i g^i \right)   
=\int \mathrm{d}^3z f \left(\partial_i \left( g^i \sqrt{\det \gamma} U'
\right) - g^i \sqrt{\det \gamma} \partial_i U\right)
\text{.} 
\end{align}
Putting these results into \eqref{FGbracket-cov}, and using
\eqref{R0tilde_def}, we get, 
\begin{align} \label{FGbracket-R0-junk}
\{F,G \} =& \int \mathrm{d}^3z \bigg[ -f \partial_i \left( g^i
  \tilde{R}^0 \right) + f g^i R_j^{(g)} \nabla_i n^j - f^i R_j^{(g)}
  \nabla_i g^j \notag \\ 
&+ 2m^4 f \sqrt{\det \gamma} \left( g^i \partial_i U - 2
  \frac{\partial U}{\partial \gamma_{mn}} \gamma_{nj} \nabla_m g^j - 2
  \frac{\partial U}{\partial \f_{mn}} \f_{nj} \bar{\nabla}_m g^j
  \right) \bigg] \text{.} 
\end{align}
The first term in this expression is the one that appears in the
covariance algebra. Thus, to obtain the correct algebra, all other
terms must cancel.  Note that, since $f^i = f n^i$, we can write,   
\begin{align}
f g^i R_j^{(g)} \nabla_i n^j - f^i R_j^{(g)} \nabla_i g^j = f g^i
R_j^{(g)} \partial_i n^j - f^i R_j^{(g)} \partial_i g^j \text{.} 
\end{align}
In the second line, writing $U=\beta_1 [U_1]+\beta_2 [U_2]+\beta_3
[U_3]$, the derivative $\dfrac{\partial U}{\partial \gamma_{mn}}$ was
computed in equations \eqref{U1}-\eqref{U3}. Now we evaluate
$\dfrac{\partial U}{\partial \f_{mn}}$ in a similar way. The
$\beta_1$-term gives, 
\begin{equation}\label{Uf1}
\beta_1\frac{\partial [U_1]}{\partial \f_{mn}}=\beta_1 \frac{\partial
  \sqrt{x}}{\partial \f_{mn}} = - \frac{\beta_1}{2} \frac{n^m
  n^n}{\sqrt{x}} \text{.} 
\end{equation}
The $\beta_2$-term, on using equation \eqref{dD1},  becomes
\begin{align} \label{Uf2}
\beta_2\frac{\partial [U_2]}{\partial \f_{mn}} 
= \beta_2 & \left( - \frac{n^m n^n}{2} {D^i}_i + n^{(m} {D^{n)}}_k n^k
+ \frac{1}{2} Q_1^{mn} \right) \text{.} 
\end{align}
The $\beta_3$-term, on using  \eqref{dD1} and \eqref{dD2}, becomes,
\begin{align}
\beta_3\frac{\partial [U_3]}{\partial \f_{mn}}
&= \beta_3 \sqrt{x}  \Bigg[\frac{1}{\sqrt{x}} \frac{\partial
\left(\sqrt{x} {D^l}_l\right)}{\partial \f_{mn}} n^i \f_{ij}{D^j}_k n^k  
-{D^i}_k n^k \f_{ij}\frac{\partial\left({D^j}_l n^l\right)}{\partial\f_{mn}} 
+ \frac{n^m n^n}{2x}{D^i}_k n^k \f_{ij} {D^j}_l n^l \notag \\ 
&+ {D^l}_l n^{(m} {D^{n)}}_k n^k - {D^m}_k n^k
  {D^n}_l n^l - \frac{n^m n^n}{2}\, e_2(D)
+ \frac{1}{2}{D^l}_l Q_1^{mn}-\frac{1}{2} Q_2^{mn} \Bigg] \text{.} 
\end{align}
Using \eqref{dD1}, the first two terms of this can be rewritten as, 
\begin{align}
&\frac{\partial \left( \sqrt{x} {D^l}_l \right)}{\partial \f_{mn}} n^i
  \f_{ij} {D^j}_k n^k - \sqrt{x} {D^i}_k n^k \f_{ij} \frac{\partial
    \left({D^j}_l n^l \right)}{\partial \f_{mn}} \notag \\ 
&\qquad\qquad= - n^l \f_{li} \left( \sqrt{x} {D^i}_j + \frac{1}{\sqrt{x}} {D^i}_r
n^r n^q \f_{qj} \right) \frac{\partial \left({D^j}_k n^k
  \right)}{\partial \f_{mn}} + \frac{1}{2} \frac{1}{\sqrt{x}} n^i
\f_{ij} {D^j}_k n^k Q_1^{mn} \text{,} 
\end{align}
When combined with the relation\footnote{Equation
  \eqref{simp_f} is analogous to equation 4.6 in \cite{fra}, and is
  derived in a similar way.} 
\begin{align} \label{simp_f}
& n^l \f_{li} \left( \sqrt{x} {D^i}_j + \frac{1}{\sqrt{x}} {D^i}_r n^r
  n^q \f_{qj} \right) \frac{\partial \left({D^j}_k n^k
    \right)}{\partial \f_{mn}} \notag \\ 
&\qquad\qquad\quad = \frac{1}{2} \frac{n^m n^n}{\sqrt{x}} {D^i}_k n^k \f_{ij} {D^j}_l
  n^l + \sqrt{x} n^{(m} {D^{n)}}_l {D^l}_k n^k - \frac{1}{2} \sqrt{x}
  {D^m}_k n^k {D^n}_l n^l \text{,} 
\end{align}
this implies that the $\beta_3$-term becomes,
\begin{align} \label{Uf3}
\beta_3\frac{\partial [U_3]}{\partial \f_{mn}}= &\beta_3 \sqrt{x} 
\bigg[{D^l}_l n^{(m} {D^{n)}}_k n^k -n^{(m} {D^{n)}}_l {D^l}_k n^k -
  \frac{1}{2}{D^m}_k n^k {D^n}_l n^l \notag \\ 
& - \frac{n^m n^n}{4}\left( {D^i}_i {D^j}_j - {D^i}_j {D^j}_i \right) 
+ \frac{1}{2}{D^l}_l Q_1^{mn} - \frac{1}{2} Q_2^{mn}
+\frac{1}{2x} n^i \f_{ij} {D^j}_k n^k Q_1^{mn} \bigg] \text{.}
\end{align}
Putting these together gives $\partial U/\partial \f_{mn}$. 

Now, observe that the last two terms of \eqref{FGbracket-R0-junk} can
be written as, 
\begin{align}
& 2 \frac{\partial U}{\partial \gamma_{mn}} \gamma_{nj} \nabla_m g^j +
  2 \frac{\partial U}{\partial \f_{mn}} \f_{nj} \bar{\nabla}_m g^j
  \notag \\ 
 &\qquad\qquad= \left( 2 \frac{\partial U}{\partial \gamma_{mn}} \gamma_{nj} + 2
 \frac{\partial U}{\partial \f_{mn}} \f_{nj} \right) \partial_m g^j  +
 2 \frac{\partial U}{\partial \gamma_{mn}} \gamma_{nj} \Gamma_{mi}^j
 g^i + 2 \frac{\partial U}{\partial \f_{mn}} \f_{nj}
 \bar{\Gamma}_{mi}^j g^i 
\end{align}
The quantity in the parenthesis is computed using
\eqref{U1}-\eqref{U3} and \eqref{Uf1}-\eqref{Uf3}. Its $\beta_1$-term
is, 
\begin{equation} \label{beta1term}
\left( 2 \frac{\partial U}{\partial \gamma_{mn}} \gamma_{nj} + 2
\frac{\partial U}{\partial \f_{mn}} \f_{nj} \right)_{\beta_1} = -
\beta_1 n^m \frac{n^l \f_{li}}{\sqrt{x}} {\delta^i}_j \text{.} 
\end{equation}
After a short calculation, using \eqref{Q1def}, the $\beta_2$-terms
become, 
\begin{align} \label{beta2term}
\left( 2 \frac{\partial U}{\partial \gamma_{mn}} \gamma_{nj} + 2
\frac{\partial U}{\partial \f_{mn}} \f_{nj} \right)_{\beta_2} =
-\beta_2 n^m \frac{n^l \f_{li}}{\sqrt{x}} \sqrt{x} \left( {\delta^i}_j
{D^k}_k - {D^i}_j \right) \text{.} 
\end{align}
Finally, the $\beta_3$-term can be written as,
\begin{align} \label{beta3term}
& \left( 2 \frac{\partial U}{\partial \gamma_{mn}} \gamma_{nj} + 2
  \frac{\partial U}{\partial \f_{mn}} \f_{nj} \right)_{\beta_3} \notag
  \\ 
&\qquad\qquad= -\beta_3 n^m \frac{n^l \f_{li}}{\sqrt{x}} x \left(
  \frac{1}{2} {\delta^i}_j \left( {D^r}_r {D^k}_k - {D^r}_k {D^k}_r
  \right) + {D^i}_k {D^k}_j - {D^i}_j {D^k}_k \right) + \beta_3 A^m_j
  \text{,} 
\end{align}
where $A^m_j$ are given by,
\begin{align}
A^m_j =&
\sqrt{x}\bigg[{D^l}_l(n^n {D^m}_k n^k + Q_1^{mn})
-(n^n {D^m}_l -\frac{1}{x}  n^i \f_{il} Q_1^{mn})   
{D^l}_k n^k- {D^m}_k n^k {D^n}_l n^l  -Q_2^{mn} 
\bigg] \f_{nj} \notag \\ 
&+\sqrt{x}\Bigg[{D^l}_l \f_{ir} { \left(D^{-1} \right)^r}_k
-  \f_{ki} +  \frac{1}{x} n^l \f_{lq} {D^q}_r n^r \f_{is} {
    \left(D^{-1} \right)^s}_k-\frac{1}{x} n^r \f_{ri} n^l \f_{lk}\Bigg]
  \frac{\partial \gamma^{ki}}{\partial \gamma_{mn}}\gamma_{nj} 
\text{.}
\end{align}
Using \eqref{Q1def} and \eqref{Q2def}, this reduces to,
\begin{align}
A^m_j= -\left(
\frac{1}{\sqrt{x}} n^s \f_{sq} {D^q}_r n^r {D^m}_i n^i 
+ \sqrt{x} {D^m}_l {D^l}_k n^k \right) n^n \f_{nj}  -
\frac{1}{\sqrt{x}} n^i \f_{ir} n^l \f_{lk} \frac{\partial
  \gamma^{rk}}{\partial \gamma_{mn}} \gamma_{nj} =0\text{.} 
\end{align}
The vanishing of $A^m_j$ follows from the definition of the matrix $D$ 
in \eqref{Ddef}, which implies,
\begin{equation} \label{D2eq}
x {D^m}_l {D^l}_k = \gamma^{ml} \f_{lk} - {D^m}_i n^i {D^l}_j n^j
\f_{lk} \text{,} 
\end{equation}
Putting \eqref{beta1term}, \eqref{beta2term} and \eqref{beta3term}
together, we get, 
\begin{align}
2m^4 f &\sqrt{\det \gamma} \left( - 2 \frac{\partial U}{\partial
    \gamma_{mn}} \gamma_{nj} - 2 \frac{\partial U}{\partial \f_{mn}}
  \f_{nj} \right) \partial_m g^j \notag \\ 
=&  2m^4 f^m \sqrt{\det \gamma} \frac{n^l \f_{li}}{\sqrt{x}} \Bigg[
    \beta_1 {\delta^i}_j + \beta_2 \sqrt{x} \left( {\delta^i}_j
         {D^k}_k - {D^i}_j \right) \notag \\ 
& + \beta_3 x \left( \frac{1}{2} {\delta^i}_j \left( {D^r}_r {D^k}_k -
         {D^r}_k {D^k}_r \right) + {D^i}_k {D^k}_j - {D^i}_j {D^k}_k
         \right) \Bigg] \partial_m g^j 
= f^i R_j^{(g)} \partial_i g^j \text{,}
\end{align}
where \eqref{biC_i} and \eqref{neqom} have been used in the last
step. This means that \eqref{FGbracket-R0-junk} simplifies to 
\begin{align} \label{FGbracket-R0-lessjunk}
\{F,G \} = \int & \mathrm{d}^3z \bigg[  -f \partial_i \left( g^i
  \tilde{R}^0 \right) + f g^i R_j^{(g)} \partial_i n^j \notag \\ 
& + 2m^4 f \sqrt{\det \gamma} \left( g^i  \partial_i U - 2
  \frac{\partial U}{\partial \gamma_{mn}} \gamma_{nj} \Gamma_{mi}^j
  g^i - 2 \frac{\partial U}{\partial \f_{mn}} \f_{nj}
  \bar{\Gamma}_{mi}^j g^i \right) \bigg] \text{.} 
\end{align}
The final two terms of this can be rewritten as,
\begin{equation}
2 \frac{\partial U}{\partial \gamma_{mn}} \gamma_{nj} \Gamma_{mi}^j
g^i = g^i \frac{\partial U}{\partial \gamma_{mn}} \partial_i
\gamma_{mn} \text{,} 
\qquad
2 \frac{\partial U}{\partial \f_{mn}} \f_{nj} \bar{\Gamma}_{mi}^j g^i
= g^i \frac{\partial U}{\partial \f_{mn}} \partial_i \f_{mn} \text{.} 
\end{equation}
Furthermore, the derivative of $U$ can be expressed as,
\begin{equation}
\partial_i U = \frac{\partial U}{\partial n^j} \partial_i n^j +
\frac{\partial U}{\partial \gamma_{mn}} \partial_i \gamma_{mn} +
\frac{\partial U}{\partial \f_{mn}} \partial_i \f_{mn} \text{.} 
\end{equation}
Note that the last two terms in this expression cancels against the
terms involving Christoffel symbols in
\eqref{FGbracket-R0-lessjunk}. Finally, from \eqref{dR0/dn}, it
follows that, 
\begin{align}
2m^4 \sqrt{\det \gamma} f g^i \frac{\partial U}{\partial n^j}
\partial_i n^j = - f g^i R_j^{(g)} \partial_i n^j \text{,} 
\end{align}
which cancels against the second term in
\eqref{FGbracket-R0-lessjunk}. The bracket
\eqref{FGbracket-R0-lessjunk} therefore becomes 
\begin{equation} \label{FGbracket-R0}
\{F,G \} = -\int \mathrm{d}^3z f \partial_i \left( g^i \tilde{R}^0 \right) \text{.}
\end{equation}
In view of \eqref{FGbracket}, this yields bracket quoted in \eqref{R0Ri-tilde},
\begin{equation}
\{ \tilde{R}^0(x),\tilde{R}_i(y) \} = - \tilde{R}^0(y)
\frac{\partial}{\partial x^i} \delta^3(x-y) \text{.} 
\end{equation}

\subsection{Evaluation of $\{ \tilde{R}^0(x),\tilde{R}^0(y) \}$}

In analogy with $F$, $B_F$ and $D_F$ defined as in \eqref{FGdef} and
\eqref{BFDF}, let us also define, 
\begin{equation}
K \equiv \int \mathrm{d}^3y k(y) \tilde{R}^0(y)\,,\qquad
B_K \equiv \int \mathrm{d}^3y k(y) R^{0 (f)}(y) \,, \qquad 
D_K \equiv \int \mathrm{d}^3y k^i(y) R_i^{(g)}(y) \text{.}
\end{equation}
where, $k^i = k n^i$ and, as in the previous section, $n^i$ can be
kept fixed when computing the variations. From this, it follows that,
\begin{equation} \label{FKbracket}
\{F,K\} = \int \mathrm{d}^3x \int \mathrm{d}^3y f(x) k(y) \{
\tilde{R}^0(x),\tilde{R}^0(y) \}\text{.} 
\end{equation}
The variations of $F$ are once again given by \eqref{dF/dgamma},
and \eqref{dF/df}, while the variations of $K$ are given by similar
equations with smoothing function $f$ replaced by $k$. 
As before, the bracket $\{F,K \}$ can be split in two parts,
$\{F,K \} = \{F,K \}_g + \{F,K \}_f$, the first of these is given by,  
\begin{align}
\{F,K \}_g  = \{D_F,D_K \}_g + 2m^4 \int \mathrm{d}^3z  \frac{\partial
  \left( \sqrt{ \det \gamma} U' \right) }{\partial \gamma_{mn}} \left(
f \frac{\delta D_K}{\delta \pi^{mn}} - k \frac{\delta D_F}{\delta
  \pi^{mn}} \right)  \text{.} 
\end{align}
The variations ${\delta D_F}/{\delta \pi^{mn}}$ and $\delta D_K/\delta
\pi^{mn}$ are given by \eqref{dA/dpi}, and the bracket $\{D_F,D_K
\}_g$ is similar to \eqref{ABbracket_g}. Putting all this together
gives,  
\begin{align}
\{F,K \}_g = \int \mathrm{d}^3z 
\bigg[2m^4\sqrt{\det \gamma}U^{mn}\gamma_{nj}\left(k^j 
\partial_m f - f^j \partial_m k \right) 
-\left(f^i R_j^{(g)}\partial_i k^j-k^j R_i^{(g)}\partial_j f^i\right)   
\bigg] \text{,}
\end{align}
where the covariant derivatives have been replaced with ordinary ones
since the terms involving Christoffel symbols cancel. 

Now we turn our attention to $\{F,K\}_f$. This bracket is given by,
\begin{align}
\{F,K\}_f = \{B_F,B_K\}_f + 2m^4 \int \mathrm{d}^3z  \sqrt{\det
  \gamma} \frac{\partial U'}{\partial \f_{mn}} \left( f \frac{\delta
  B_K}{\delta p^{mn}} - k \frac{\delta B_F}{\delta p^{mn}}  \right)
\text{.} 
\end{align}
The variations of $B_F$ and $B_K$ with respect to $p^{ij}$ are given
by \eqref{dAH/dpi} and the terms under the integral sign cancel out.
The bracket $\{B_F,B_K\}$ is given by \eqref{AHBHbracket_g}, now with 
$R^{i (f)} \equiv \f^{ij}R_j^{(f)}$. Hence,  
\begin{align}
\{F,K\}_f =  \{B_F,B_K\}_f =  - \int \mathrm{d}^3z \left( f R^{i (f)} \partial_i k - k R^{i (f)} \partial_i f \right) \text{,}
\end{align}
Combining $\{F,K\}_f$ and $\{F,K\}_g$ yields the expression
\begin{align} \label{FKsimp}
\{F,K\} = \int \mathrm{d}^3z \bigg[-R^{i (f)}\left( f\partial_i k 
- k\partial_i f \right) 
& - R_i^{(g)}\left(f^j\partial_j k^i-k^j\partial_j f^i\right) 
\notag \\ &
+ 2m^4 \sqrt{\det \gamma} U^{mn} \gamma_{nj} \left( k^j \partial_m f -
  f^j \partial_m k \right) \bigg] \text{.}
\end{align}
Since $k^i=n^ik$ and $f^i=n^if$, the second term in this expression is,
\begin{align} \label{fkR-rewrite}
- R_i^{(g)} \left( f^j\partial_j k^i - k^j\partial_j f^i \right)=  n^i
R_i^{(g)} \left( k^j \partial_j f - f^j \partial_j k \right) \text{.} 
\end{align}
By the definition of $U^{mn}$ \eqref{Umn-def}, the last term in
\eqref{FKsimp} is,
\begin{align}
 2 m^4 \sqrt{\det \gamma} &U^{mn} \gamma_{nj} \left( k^j \partial_m f
 - f^j \partial_m k \right) \notag \\ 
 =  2 & m^4 \sqrt{\det \gamma} U \left( k^j \partial_j f - f^j
 \partial_j k \right) + 4m^4 \sqrt{\det \gamma} \frac{\partial
   U}{\partial \gamma_{mn}} \gamma_{nj} \left( k^j \partial_m f - f^j
 \partial_m k \right)\,. 
\end{align}
Then, using the expression for $U$ in \eqref{CDstep1} we have,
\begin{align}
2 m^4\sqrt{\det \gamma}U\left(k^j\partial_j f-f^j \partial_j k\right) 
= 2m^4 \sqrt{\det \gamma} &\left[\frac{\beta_1}{\sqrt{x}} 
+\beta_2 {D^l}_l + \beta_3\sqrt{x} e_2(D) \right] 
\left( k^j \partial_j f - f^j \partial_j k \right) 
\notag\\
&\qquad\,\, - n^i R_i^{(g)} \left( k^j \partial_j f - f^j \partial_j k \right) \text{.}
\end{align}
The last term cancels against the expression in \eqref{fkR-rewrite},
so that \eqref{FKsimp} becomes, 
\begin{align} \label{FKsimp2}
\{F,K\} = & \int \mathrm{d}^3z \Bigg[ - \left( f R^{i (f)} \partial_i
  k - k R^{i (f)} \partial_i f \right)  \notag \\ 
 & + 2 m^4 \sqrt{\det \gamma} \left( \frac{\beta_1}{\sqrt{x}} +
  \beta_2 {D^l}_l + \frac{\beta_3}{2} \sqrt{x} \left( {D^k}_k {D^r}_r
  - {D^k}_r {D^r}_k \right)  \right) \left( k^j \partial_j f - f^j
  \partial_j k \right) \notag \\ 
& + 4 m^4 \sqrt{\det \gamma} \frac{\partial U}{\partial \gamma_{mn}}
  \gamma_{nj} \left( k^j \partial_m f - f^j \partial_m k \right)
  \Bigg] \text{.} 
\end{align}
We now concentrate on the last term of \eqref{FKsimp2}. The
$\beta_2$-part of this, using equation \eqref{U2}, becomes, 
\begin{align} \label{beta2calc}
\left[ 4 m^4 \sqrt{\det \gamma} \frac{\partial U}{\partial
    \gamma_{mn}} \gamma_{nj} \left( k^j \partial_m f - f^j \partial_m
  k \right) \right]_{\beta_2} 
 = -  2  m^4 \sqrt{\det \gamma} \beta_2  {D^m}_j \left( k^j \partial_m
 f - f^j \partial_m k \right)\,, 
\end{align}
where we have used \eqref{xdef}, together with the relation
\begin{equation} \label{Deq}
\f_{nk} {\left(D^{-1} \right)^k}_l \gamma^{ml} = x {D^m}_n + {D^m}_i
n^i n^k \f_{kn} \text{,} 
\end{equation}
which follows from \eqref{Ddef}. The $\beta_3$-part takes the form, on
using \eqref{U3},
\begin{align}
& \left[ 4 m^4 \sqrt{\det \gamma} \frac{\partial U}{\partial
\gamma_{mn}} \gamma_{nj} \left( k^j \partial_m f - f^j
\partial_m k \right) \right]_{\beta_3} 
=2 m^4 \sqrt{\det \gamma} \beta_3 \Bigg[  \sqrt{x} {D^l}_l 
\f_{ir} {\left(D^{-1} \right)^r}_k -  \sqrt{x} \f_{ik} 
\notag \\
&\qquad\qquad\quad + \frac{1}{\sqrt{x}}  n^l \f_{lq} {D^q}_r n^r 
\f_{is} {\left(D^{-1} \right)^s}_k - \frac{1}{\sqrt{x}} n^r \f_{ri} 
n^l \f_{lk}\Bigg] \frac{\partial \gamma^{ik}}{\partial \gamma_{mn}}   
\gamma_{nj}\left( k^j\partial_m f - f^j \partial_m k \right)\text{.}
\end{align}
Using \eqref{xdef}, \eqref{D2eq} and \eqref{Deq}, many of these terms
cancel in a way similar to the $\beta_2$ case in
\eqref{beta2calc}. After some calculations, the $\beta_3$-part reduces
to 
\begin{align}
& \left[ 4 m^4 \sqrt{\det \gamma} \frac{\partial U}{\partial
    \gamma_{mn}} \gamma_{nj} \left( k^j \partial_m f - f^j \partial_m
  k \right) \right]_{\beta_3} \notag \\
&\qquad\qquad\qquad\qquad = 2 m^4 \sqrt{\det \gamma} \beta_3 \sqrt{x} \left(
  {D^m}_k {D^k}_j - {D^m}_j {D^k}_k \right) \left( k^j \partial_m f -
  f^j \partial_m k \right) \text{.} 
\end{align}
Finally, combining these results with equation \eqref{neqom}, and
performing some manipulations, allows the bracket \eqref{FKsimp2} to
be expressed as, 
\begin{align}
\{F,K\} &= \int \mathrm{d}^3z \bigg[ - \left( f R^{i (f)} \partial_i k
  - k R^{i (f)} \partial_i f \right) - \f^{ij} R_j^{(g)} \left( f
  \partial_i k - k \partial_i f \right) \bigg] \notag \\ 
&= - \int \mathrm{d}^3z \left( f \tilde{R}^i \partial_i k - k
\tilde{R}^i \partial_i f \right) \text{,} 
\end{align}
where we have defined
\begin{equation}
\tilde{R}^i \equiv \f^{ij} \tilde{R}_j  = \f^{ij} R_j^{(g)} + R^{i (f)} \text{.}
\end{equation}
Combining this with \eqref{FKbracket} yields the bracket \eqref{R0R0-tilde},
\begin{equation}
\{ \tilde{R}^0(x),\tilde{R}^0(y) \} = - \left( \tilde{R}^i(x) \frac{\partial}{\partial x^i} \delta^3(x-y) - \tilde{R}^i(y) \frac{\partial}{\partial y^i} \delta^3(x-y) \right) \text{,}
\end{equation}

\subsection{Poisson brackets with the constraint $\mathcal{C}$} \label{Cbracketsection}

We will now compute the Poisson brackets of $\tilde{R}^0$ and
$\tilde{R}_i$ with $\mathcal{C}$ in \eqref{bi-C}. Let us define,
\begin{equation}
F_C \equiv \int \mathrm{d}^3x f(x) \mathcal{C}(x) \text{.}
\end{equation}
The variations of $F_C$ with respect to $\gamma_{ij}$, $\pi^{ij}$,
$\f_{ij}$ and $p^{ij}$ are,
\begin{align} \label{dFC/dgamma}
\frac{\delta F_C}{\delta \gamma_{mn}} &= \frac{\delta A^0}{\delta
  \gamma_{mn}} + \frac{\delta A}{\delta \gamma_{mn}} + 2m^4 f
\frac{\partial \left( \sqrt{ \det \gamma} V \right)}{\partial
  \gamma_{mn}} \text{,} 
\qquad
\frac{\delta F_C}{\delta \pi^{mn}} = \frac{\delta A^0}{\delta\pi^{mn}}  
+ \frac{\delta A}{\delta \pi^{mn}} \text{,} \\
\label{dFC/df}
\frac{\delta F_C}{\delta \f_{mn}} &=  \frac{\delta A}{\delta \f_{mn}} +
2m^4 f \sqrt{ \det \gamma} \frac{\partial  V}{\partial \f_{mn}} \text{,}
\qquad\qquad\quad\,\,\,\,\,
\frac{\delta F_C}{\delta p^{mn}} = 0 \text{,}
\end{align}
where,
\begin{equation}
A^0 \equiv \int \mathrm{d}^3x f(x) R^{0 (g)}(x) \text{,} \qquad A
\equiv \int \mathrm{d}^3x f(x) R_i^{(g)}(x) {D^i}_j(x) n^j(x) \text{.} 
\end{equation}
Now, for $G$ defined in \eqref{FGdef}, we evaluate the bracket $\{F_C,G \}$
by spliting it in two parts,  
\begin{equation}
\{F_C,G \} = \{F_C,G \}_g + \{F_C,G \}_f \text{.}
\end{equation}
The $g$-bracket is given by,
\begin{align}
\{F_C,G \}_g &=  \{A^0,G \}_g + \{A,G \}_g + 2m^4 \int \mathrm{d}^3z f
\frac{\partial\left(\sqrt{\det\gamma} V \right)}{\partial\gamma_{mn}}  
\frac{\delta G}{\delta \pi^{mn}} \text{.} 
\end{align}
The first bracket in this expression is similar to
\eqref{AHBbracket_g}, and therefore becomes 
\begin{equation}
 \{A^0,G \}_g = - \int \mathrm{d}^3z f \nabla_j \left(g^j R^{0 (g)} \right) 
=- \int \mathrm{d}^3z f \partial_i \left(g^i R^{0 (g)} \right)
\end{equation}
The last step holds since $g^i$ is a vector and $R^{0 (g)}$ is a
scalar density, in analogy with \eqref{densityderivative}.  The second
bracket is similar to \eqref{ABbracket_g}, but there will be an extra
term due to the fact that the variation of ${D^i}_j n^j$ with respect
to $\gamma_{ij}$ does not vanish. The bracket therefore becomes
\begin{align}
\{A,G \}_g =-\int \mathrm{d}^3z \left(f{D^i}_j n^j  R_k^{(g)}\nabla_i g^k 
- g^k R_i^{(g)} \nabla_k \left(f {D^i}_j n^j \right)  
- f R_i^{(g)} \frac{\partial \left({D^i}_j n^j\right)}{\partial\gamma_{mn}} 
\frac{\delta G}{\delta \pi^{mn}} \right)\text{.}
\end{align}
The $f$-bracket $\{F_C,G \}_f$ is given by,
\begin{align}
\{F_C,G \}_f  = \int \mathrm{d}^3z \Big[ f R_i^{(g)} \frac{\partial \left({D^i}_j n^j \right)}{\partial \f_{mn}} \frac{\delta G}{\delta p^{mn}}  + 2m^4 f \sqrt{ \det \gamma} \frac{\partial  V}{\partial \f_{mn}} \frac{\delta G}{\delta p^{mn}} \Big] \text{.}
\end{align}
Putting these together gives,
\begin{align} \label{FCGbracket}
\{F_C,G \} = \int \mathrm{d}^3z \Big[g^k R_i^{(g)} \nabla_k  \left(f {D^i}_j n^j \right)  
&- f \partial_j \left(g^j R^{0 (g)} \right)  -  f {D^i}_j n^j  R_k^{(g)} \nabla_i g^k 
\notag \\
&\qquad + f S^{mn} \frac{\delta G}{\delta \pi^{mn}} + 
f Z^{mn} \frac{\delta G}{\delta p^{mn}} \Big]\text{.}
\end{align}
Here, $S^{mn}$ is defined in \eqref{Sdef} 
and $Z^{mn}$ is defined in \eqref{Zdef}.
The first term can be rewritten as,
\begin{equation}
\int \mathrm{d}^3z g^k R_i^{(g)} \nabla_k \left(f {D^i}_j n^j \right)= 
-\int \mathrm{d}^3z f \partial_i\left(g^i R_j^{(g)} {D^j}_k n^k\right) 
+ \int \mathrm{d}^3z f g^k R_i^{(g)} \nabla_k \left( {D^i}_j n^j \right) \text{.}
\end{equation}
The expression for the variation of $G$ given in \eqref{dG/dpi},
together with \eqref{SV}, allows us to write 
\begin{equation}
f S^{mn} \frac{\delta G}{\delta \pi^{mn}}=-2m^4 f \sqrt{\det\gamma}V 
\nabla_j g^j + 2m^4 f \sqrt{\det \gamma} \bar{V}^{mn} \gamma_{nj} 
\nabla_m g^j \text{,}
\end{equation}
where the first term can be further simplified using,
\begin{align}
\sqrt{\det \gamma} V \nabla_j g^j=\partial_i \left( g^i \sqrt{\det \gamma} V \right) 
-  g^i  \sqrt{\det \gamma} \partial_i V \text{.} 
\end{align}
In a similar way, using \eqref{ZW} and \eqref{dG/dp}, we can write
\begin{equation}
f Z^{mn} \frac{\delta G}{\delta p^{mn}} = -2m^4 f \sqrt{\det \gamma}
\bar{W}^{mn} \f_{nj} \bar{\nabla}_m g^j \text{.}
\end{equation}
Putting all of this together, the bracket $\{F_C,G \}$ becomes
\begin{align} \label{FCGbracket-C}
\{F_C,G \} = \int \mathrm{d}^3z f \left[-\partial_i \left( g^i
  \mathcal{C} \right) + \Delta' \right]\text{.} 
\end{align}
where $\Delta'$ is the same as the quantity $\Delta$ in \eqref{Delta}
with $L^i$ replaced by $g^i$,
\begin{align} \label{Delta'}
\Delta' &=
R_j^{(g)}\left(\nabla_i({D^j}_k n^k)g^i - {D^i}_k n^k\nabla_i g^j\right) 
\notag\\ 
& \qquad\qquad\qquad
+2m^4\sqrt{\det\gamma}
\left( \partial_i V g^i+ \gamma_{jk} \bar{V}^{ki} \nabla_i g^j 
 -\f_{jk} \bar{W}^{ki} \bar{\nabla}_i g^j \right)  \text{.}
\end{align}
We now show that $\Delta'=0$. First, note that the Christoffel symbols
in the first two terms cancel, 
\begin{equation}\label{noGamma}
R_j^{(g)}\left(\nabla_i({D^j}_k n^k)g^i - {D^i}_k n^k\nabla_i
g^j\right) =
R_j^{(g)}\left(\partial_i ({D^j}_k n^k)g^i - {D^i}_k n^k\partial_i
g^j\right) 
\end{equation}
The last two terms of $\Delta'$ are,
\begin{align} \label{cov-terms}
& \gamma_{jk} \bar{V}^{ki} \nabla_i g^j -  \f_{jk} \bar{W}^{ki}
  \bar{\nabla}_i g^j   =  \left( \bar{V}^{ki} \gamma_{jk}  -
  \bar{W}^{ki} \f_{jk} \right) \partial_i g^j +  \left( \bar{V}^{ki}
  \gamma_{jk} \Gamma^j_{il} -  \bar{W}^{ki} \f_{jk}
  \bar{\Gamma}^j_{il} \right) g^l \text{.} 
\end{align}
Using the symmetry relation \eqref{Dfsymmetry}, it follows from
\eqref{Vbardef} and \eqref{WQ} that 
\begin{align}
\bar{V}^{ki} \gamma_{jk}  - \bar{W}^{ki} \f_{jk} &= \frac{n^k
  \f_{km}}{\sqrt{x}} {D^i}_l n^l \Bigg[ \beta_1 {\delta^m}_j + \beta_2
  \sqrt{x} \left( {\delta^m}_j {D^r}_r - {D^m}_j \right) \notag \\ 
&+ \beta_3 x \left( \frac{1}{2} {\delta^m}_j \left( {D^r}_r {D^h}_h -
       {D^r}_h {D^h}_r \right) + {D^m}_n {D^n}_j - {D^m}_j {D^r}_r
       \right) \Bigg] \text{,} 
\end{align}
which, together with \eqref{biC_i} and \eqref{neqom}, implies that
\begin{equation} \label{VWR}
2  m^4 \sqrt{\det \gamma} \left( \bar{V}^{ki} \gamma_{jk}  -
\bar{W}^{ki} \f_{jk} \right) = R_j^{(g)} {D^i}_k n^k \text{.} 
\end{equation}
Also, since $\bar{V}^{ki}$ and $\bar{W}^{ki}$ are symmetric, the last
two terms in \eqref{cov-terms} become,
\begin{equation} \label{sym_Chris}
\bar{V}^{ki} \gamma_{jk} \Gamma^j_{il} = \frac{1}{2} \bar{V}^{ki}
\partial_l \gamma_{ik} 
\text{,} \qquad 
\bar{W}^{ki} \f_{jk}\bar{\Gamma}^j_{il} = \frac{1}{2} \bar{W}^{ki} 
\partial_l \f_{ik}
\text{.} 
\end{equation}
The above equations can be used to rewrite $\Delta'$ as,
\begin{align} \label{Delta'rewrite1}
\Delta' &=
R_j^{(g)}\partial_i({D^j}_k n^k)g^i  
+m^4\sqrt{\det\gamma}\left[ 2\partial_i V g^i+ 
\left(\bar{V}^{mn} \partial_i \gamma_{mn} 
-\bar{W}^{mn} \partial_i \f_{mn}\right) g^i \right] \text{.}
\end{align}
We will now compute $\partial_i V$. From \eqref{Vdef} it follows that,  
\begin{align} \label{dV/dx}
\partial_i V &= \beta_1 \partial_i \left( \sqrt{x} {D^j}_j \right) + \frac{1}{2} \beta_2 \Big(2 \sqrt{x} {D^j}_j \partial_i \left( \sqrt{x} {D^k}_k \right) - \partial_i \left( x {D^j}_k {D^k}_j \right) \Big) \notag \\
&+\frac{1}{6} \beta_3 \bigg( 3x {D^j}_j {D^k}_k \partial_i \left( \sqrt{x} {D^l}_l \right) - 3x {D^k}_l {D^l}_k \partial_i \left( \sqrt{x} {D^j}_j \right) \notag \\
&- 3 \sqrt{x} {D^j}_j \partial_i \left( x {D^k}_l {D^l}_k \right) + 2  \partial_i \left( x^{3/2} {D^j}_k {D^k}_l {D^l}_j \right) \bigg) \text{.}
\end{align}
To evaluate this, we need the following derivatives that can be
computed using equation \eqref{Ddef}, 
\begin{align}
\partial_i \left( \sqrt{x} {D^j}_j \right) &= \frac{1}{2} \frac{1}{\sqrt{x}} \left[ -2 n^j \f_{lj} \partial_i \left({D^l}_n n^n \right) - \f_{lj} {\left(D^{-1} \right)^j}_k \gamma^{km} \gamma^{ln} \partial_i \gamma_{mn} + Q_1^{jl} \partial_i \f_{lj} \right] \text{,} \notag \\
\partial_i \left( x {D^j}_k {D^k}_j \right) &=  -2 {D^j}_m n^m \f_{lj} \partial_i \left({D^l}_n n^n \right) - \f_{lj} \gamma^{jm} \gamma^{ln} \partial_i \gamma_{mn} + Q_2^{jl} \partial_i \f_{lj} \text{,} \\
\partial_i \left( x^{3/2} {D^j}_k {D^k}_l {D^l}_j \right) &= \frac{3}{2} \sqrt{x} \left[ -2 {D^j}_k {D^k}_m n^m \f_{lj} \partial_i \left({D^l}_n n^n \right) - \f_{lj} {D^j}_k \gamma^{km} \gamma^{ln} \partial_i \gamma_{mn} + Q_3^{jl} \partial_i \f_{lj} \right] \text{.} \notag
\end{align}
Using these expressions, as well as \eqref{Vbardef} and \eqref{WQ},
equation \eqref{dV/dx} can be rewritten as 
\begin{align}
\partial_i V &= - \Bigg[ \frac{\beta_1}{\sqrt{x}} n^j \f_{lj} + \beta_2 \left( {D^r}_r n^j \f_{lj} - {D^j}_m n^m \f_{lj} \right) \notag \\
& + \beta_3 \sqrt{x} \left( \frac{1}{2} \left( {D^r}_r {D^h}_h - {D^r}_h {D^h}_r \right) n^j \f_{lj} - {D^r}_r {D^j}_m n^m \f_{lj} + {D^j}_k {D^k}_m n^m \f_{lj} \right) \Bigg] \partial_i \left( {D^l}_n n^n \right) \notag \\
& - \frac{1}{2} \bar{V}^{mn} \partial_i \gamma_{mn} + \frac{1}{2} \bar{W}^{mn} \partial_i \f_{mn} \text{,}
\end{align}
which, together with \eqref{Dfsymmetry} and \eqref{biC_i} implies
\begin{equation} \label{dV}
2  m^4 \sqrt{\det \gamma} \partial_i V = 
-R_j^{(g)} \partial_i \left({D^j}_k n^k \right) - m^4 \sqrt{\det \gamma} \left( \bar{V}^{mn}
\partial_i \gamma_{mn} - \bar{W}^{mn} \partial_i \f_{mn} \right)
\text{.} 
\end{equation}
Substituting this in \eqref{Delta'rewrite1} gives $\Delta'=0$, as
desired, and one obtains,
\begin{equation}
\{F_C,G \} = - \int \mathrm{d}^3z  f \partial_i \left( g^i \mathcal{C} \right) \text{.}
\end{equation}
This expression is of the same form as \eqref{FGbracket-R0}, but with $\tilde{R}^0$ replaced by $\mathcal{C}$. By the same argument as for that bracket, it follows that
\begin{equation} \label{CRibracket_app}
\{ \mathcal{C}(x),\tilde{R}_i(y) \} = - \mathcal{C}(y) \frac{\partial}{\partial x^i} \delta^3(x-y) \text{.}
\end{equation}
Since $\mathcal{C}$ is a scalar density, it should transform as such under spatial diffeomorphisms. Note that the Poisson bracket above is of the same form as \eqref{R0Ri-tilde}. Since $\tilde{R}^0$ is also a scalar density, this means that $\mathcal{C}$ transforms in the appropriate way, consistent with our earlier results.

Finally, we look at the bracket $\{\mathcal{C}(x),\tilde{R}^0(y) \}$. Recall from \eqref{C2bracket} that
\begin{align}
\mathcal{C}_{(2)}(x) \approx - \int \mathrm{d}^3y \left[ L^i(y) \{\mathcal{C}(x),  \tilde{R}_i(y) \} + L(y) \{ \mathcal{C}(x), \tilde{R}^0(y) \} \right] \text{.}
\end{align}
Imposing the constraint $\mathcal{C}=0$, equation \eqref{CRibracket_app} implies that
\begin{equation} \label{CR0bracket}
\int \mathrm{d}^3y  L(y) \{ \mathcal{C}(x), \tilde{R}^0(y) \} = - \mathcal{C}_{(2)}(x) + L^i(x) \frac{\partial}{\partial x^i} \mathcal{C}(x) \approx - \mathcal{C}_{(2)}(x) \text{,}
\end{equation}
since the spatial derivative of $\mathcal{C}$ vanishes on the
constraint surface. In particular, equation \eqref{CR0bracket} shows
that $\C$ is independent of $L^i$, as discussed in section
\ref{Li-dep}. Then it follows that computing the bracket $\{
\mathcal{C}(x), \tilde{R}^0(y) \}$ involves preforming essentially the
same calculations as those in section \ref{biseccon}. The result is
\begin{equation} \label{CR0bracket_final}
\{ \mathcal{C}(x), \tilde{R}^0(y) \} = - \frac{\C(x)}{L(x)} \delta^3(x-y) \text{.}
\end{equation}
The right hand side of \eqref{CR0bracket_final} is independent of $L$,
as can be seen from \eqref{bimetricC2_final_form}.

\bibliographystyle{JHEP}
\bibliography{referens-arXiv-v2}

\end{document}